\documentstyle{mn}

\input epsf

\begin{document}

\newcommand{\Ob}{\Omega_{\rm B}}
\def\ga{\mathrel{\lower0.6ex\hbox{$\buildrel {\textstyle >}
 \over {\scriptstyle \sim}$}}}
\def\la{\mathrel{\lower0.6ex\hbox{$\buildrel {\textstyle <}
 \over {\scriptstyle \sim}$}}}
\def\kms{\;{\rm km\,s^{-1}}}
\def\kmsmpc{\;{\rm km\,s^{-1}\,Mpc^{-1}}}
\def\hompc{\,h\,{\rm Mpc}^{-1}}
\def\mpcoh{\,h^{-1}\,{\rm Mpc}}

\def\vec#1{{\bf #1}}

\def\S{Section\ } 

\journal{\hfill\hfill
\it Submitted for publication in MNRAS
\hfill}
 
\title[Baryonic Signatures in Large-Scale Structure]{Baryonic
Signatures in Large-Scale Structure}

\author[A.Meiksin et al.]{A. Meiksin${}^{1}$,
Martin White${}^{2}$ and J.A. Peacock${}^{1}$ \\
${}^{1}$Institute for Astronomy, University of Edinburgh,
Royal Observatory, Blackford Hill, Edinburgh EH9 3HJ, UK\\
${}^{2}$Departments of Physics and Astronomy, University of Illinois
at Urbana-Champaign, Urbana, IL 61801-3080, USA}
\pubyear{1998}

\maketitle

\begin{abstract}
We investigate the consequences of a non-negligible baryon fraction
for models of structure formation in Cold Dark Matter dominated
cosmologies, emphasizing in particular the existence of oscillations
in the present-day matter power spectrum.  These oscillations are the
remnants of acoustic oscillations in the photon-baryon fluid before
last scattering, for which evidence from measurements of the Cosmic
Microwave Background anisotropy is mounting. For acceptable values of
the cosmological and baryon densities, the oscillations modulate the
power by up to $\sim 10\%$, with a `period' in spatial wavenumber
which is close to $\Delta k\simeq0.05\;{\rm Mpc}^{-1}$.
We study the effects of nonlinear evolution on these features, and show
that they are erased for $k\ga 0.2 \hompc$.
At larger scales, the features evolve as expected from second-order
perturbation theory: the visibility of the oscillations is affected
only weakly by nonlinear evolution.
No realistic CDM parameter combination is able to account for the claimed
feature at $k\simeq 0.1\mpcoh$ in the APM power spectrum, or the
excess power at $100\mpcoh$ wavelengths quoted by several recent surveys.
Thus baryonic oscillations are not predicted to dominate existing measurements
of clustering.
We examine several effects which may mask the features which {\it are\/}
predicted, and conclude that future galaxy surveys may be able to detect
the oscillatory features in the power spectrum provided baryons comprise
$\ga 15\%$ of the total density, but that it will be a technically
challenging achievement.
\end{abstract}
 

\section{Introduction}

It is commonly assumed that the mass of the Universe is dominated by
Cold Dark Matter (CDM) and that the baryons had a negligible influence
on the development of large-scale structure. Recent estimates of the
ratio of the baryon density to dark matter density suggest that this
view may be incorrect. Because of the coupling between baryons and the
Cosmic Microwave Background (CMB) radiation during the recombination
epoch, additional structure in the power spectrum will develop
compared with a standard CDM scenario in which the power spectrum
contains only one feature, at the horizon scale of matter-radiation equality.
We discuss the detectability of such higher
order effects in the context of currently favored models for structure
formation. We show that they will leave a significant imprint on the
matter power spectrum that will be measurable with forthcoming galaxy
redshift surveys for a wide range of parameters, including
the possibility of a direct detection of acoustic oscillations in the
galaxy power spectrum. Unlike the CMB, the imprint of the oscillations on
the matter power spectrum will survive even if the universe underwent
early reionization, in which case they could only be revealed by large
galaxy surveys.

The baryon density as determined by Big Bang Nucleosynthesis (BBN) has been
on the rise in recent years (see discussion in 
White et al.~\shortcite{WVLS}).
At the beginning of this decade, the baryon density was estimated to
have a 95\% confidence range of
$\Ob h^2=0.0125\pm0.0025$  \cite{Waletal,Smietal},
where $\Ob$ is the baryon density in units of the critical density, and $h$
is the present Hubble parameter in units of
100$\,{\rm km}\,{\rm s}^{-1}\,{\rm Mpc}^{-1}$.
This range has recently broadened and shifted toward higher values.  For
example, the measurement of the primordial deuterium abundance by
Tytler, Fan \& Burles \shortcite{TytFanBur} yields
$\Ob h^2=0.024\pm0.002\pm0.002\pm0.001$, with the 1$\sigma$
uncertainties being statistical, systematic, and theoretical respectively.
While extragalactic assessments of $\Ob$ have crept upwards, the measured
abundance of deuterium in the ISM sets a ceiling of $\Ob h^2<0.031$
\cite{Linetal}, within the assumptions of the standard BBN paradigm.
It may be possible to relax the nucleosynthesis bound on $\Ob$ with new
particle physics, an example being the decaying tau neutrino proposal
of Gyuk \& Turner \shortcite{GuyTur}, or by allowing inhomogeneities in the
baryon-to-photon ratio
(see e.g., Mathews, Kajino \& Orito~1996 and references therein).
However, in the latter case, even with relaxed constraints on the primordial
${}^{7}$Li abundance one finds $\Ob h^2<0.0325$.

Support for a high baryon fraction is also provided by clusters of galaxies
\cite{Whietal,WhiFab,ElbArnBoh,Maretal}.
The recent compilation by White \& Fabian \shortcite{WhiFab} gives
\begin{equation}
{\Ob\over\Omega_0} = 0.14_{-0.04}^{+0.08} \, \left(h\over0.5\right)^{-3/2}
\end{equation}
(95\% confidence) with comparable lower limits quoted by other authors:
Steigman \& Felten~\shortcite{SteFel} find
${\Ob/\Omega_0}\ge 0.2 \left(h/0.5\right)^{-3/2}$,
while Evrard, Metzler \& Navarro \shortcite{EvrMetNav} estimate
${\Ob/\Omega_0}\ge 0.11 \left(h/0.5\right)^{-3/2}$.
Here $\Omega_0\equiv\Omega_{\rm CDM}+\Ob$ is the total non-relativistic
energy density relative to the critical density.
Given the above discussion we consider $0.01\le \Ob h^2\le 0.03$ and
$0.1\le \Omega_0 h^2\le 0.25$ as a fair range of parameters and we shall
work within this range except in \S\ref{sec:clustering} where we shall relax
these assumptions.

For some time it has been recognized that a universe with a
substantial baryon fraction will result in significant features in the
matter power spectrum.  Historically, Sakharov predicted oscillations
in the matter power spectrum in a cold universe with no radiation
component, building on earlier work of Lifshitz.  The oscillations
were an imprint of sound waves as in the modern context, but the
restoring pressure was due to degenerate electron pressure at high
densities, not the CMB photons.  The first calculations to emphasize
the oscillations in the context of hot big bang CDM models with
isocurvature and adiabatic fluctuations concentrated on models with
very low $\Omega_0\sim0.1$ and high $\Ob\sim\Omega_0$
\cite{Dek,BBE,PeeA,PeeB,BDP,Sug,HuSug}.  The features are extremely
prominent in this case -- but the parameter values involved in these
models lie well outside of the currently preferred range.

In this paper, we investigate more realistic values of the parameters where
the features are correspondingly more difficult to detect.  A study of
the CDM $+$ baryon transfer function over
a wide range of the $\Omega_0$--$\Ob$ parameter space has recently
been performed by Eisenstein \& Hu~\shortcite{EisHu} and Eisenstein et
al.~\shortcite{EHSS}, and the detectability of oscillations has been
studied by Tegmark~\shortcite{Teg97} and
Goldberg \& Strauss~\shortcite{GolStr}.
Where our results overlap they are in good agreement; our work concentrates
on the role of nonlinear growth in modifying the form of the oscillations,
which has not previously been explored.

The outline of the  paper is as follows.  In \S\ref{sec:linear} we discuss
the predictions for the oscillations in the matter power spectrum from linear
theory
(some of the relevant technical details are in Appendix \ref{sec:theory}).
The effects of non-linearities are included in \S\ref{sec:nonlinear} using 2nd
order perturbation theory and N-body simulations (our implementation of a PM
code is discussed in Appendix \ref{sec:pmcode}).
In \S\ref{sec:clustering} we discuss the existing evidence for large scale
features in the power spectrum, and conclude that baryonic features are
unlikely to account for the existing power-spectrum data.
In \S\ref{sec:measure} we discuss the measurability of these features in
forthcoming redshift surveys and in larger samples selected via photometric
redshifts.  This section includes a discussion of expected error bars, plus
systematic effects which need to be considered, some of the details of which
are outlined in Appendices \ref{sec:sdbias} and \ref{sec:counts}.
We summarize our conclusions in \S\ref{sec:summary}.

\section{Linear Theory} \label{sec:linear}
 
\begin{figure}
\begin{center}
\leavevmode \epsfxsize=3.3in \epsfbox{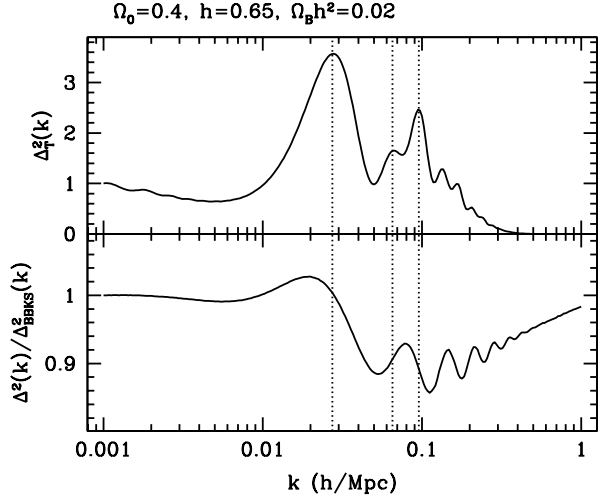}
\end{center}
\caption{A comparison of the power spectra for the CMB and LSS,
for a model with $\Omega_0=0.4$, $\Lambda=0$, $h=0.65$ and $\Ob=0.045$.
The upper panel is contribution per $\ln(k)$ 
to the total variance in the CMB (scaled to unity at the
smallest wavenumbers).
The lower panel is the power per $\ln(k)$ in the matter,
i.e.~$\Delta^2(k)$, with the trend taken out using the BBKS fitting
function (see text).
The vertical dotted lines mark the positions of the peaks in the CMB
power spectrum, which are $90^\circ$ out of phase with the 
corresponding oscillations in the LSS power spectrum.}
\label{fig:cmblss}
\end{figure}

There is considerable evidence to suggest that the matter density of
the universe is less than critical ($\Omega_0<1$). We have discussed
above the growing evidence favouring a baryon fraction larger than has
been assumed until recently.  This has motivated us to look
in detail at a variety of CDM models, including those with
$\Omega_0<1$, and to examine cases with a baryon fraction consistent
with current constraints.  For some of this regime the gravitational
potential is not totally dominated by the CDM and features in the
matter power spectrum
\begin{equation}
\Delta_{\rm L}^2(k) \equiv {k^3 P(k)\over 2\pi^2}
  = \delta_{\rm H}^2 \left( {ck\over H_0} \right)^{3+n} T^2(k)
\end{equation}
will be generated. Here $\delta_{\rm H}$ is the fractional density 
contrast $(\delta\rho/\rho)_k$ at horizon crossing $k=H_0/c$,
and $T(k)$ is the transfer function. Specifically, the linear
power spectrum $\Delta_{\rm L}^2(k)$ contains a series of peaks of small
amplitude
which arise from acoustic oscillations in the baryon-photon fluid prior to
recombination, the same source as the much larger peaks in the CMB power
spectrum, for which there is growing evidence.

We show in Fig.~\ref{fig:cmblss} the radiation and matter power spectra for
a CDM model with $\Omega_0=0.4$, $h=0.65$ and $\Ob h^2=0.02$, calculated by
numerical evolution of the coupled Einstein, Boltzmann and fluid equations.
Note that the radiation power spectrum $\Delta^2_T$ is the contribution
per $\ln k$ to the total variance in the CMB; it is closely related
to the angular power spectrum $\ell^2 C_\ell$, but the two should
not be confused.
The top panel shows the radiation power spectrum, which exhibits a clear
series of peaks. These peaks are modes which are density maxima and minima
of the oscillations of the photon-baryon fluid at recombination (see
Appendix \ref{sec:theory} for more details). The bottom panel shows the square
of the matter transfer function $T(k)$ for this same
model, with the gross features (the bend from $T(k)\sim1$ as $k\to0$
to $T(k)\sim k^{-2}$ at $k\to\infty$) removed by dividing by the fitting
function for $T(k)$ provided by Bardeen et al.~\shortcite{BBKS} (hereafter
BBKS)
\begin{equation}
\begin{array}{lcr}
T(k) &=& {\displaystyle{\ln\left(1+2.34q\right)\over 2.34q}}
   \left[1+3.89q+(16.1q)^2 \;+\right. \\
  & &\left. (5.46q)^3+(6.71q)^4\right]^{-1/4} \,,
\end{array}
\end{equation}
with $q \equiv k/h\Gamma^{*}$ and $\Gamma^{*}$ a parameter.
The BBKS fitting function provides a reasonable fit to $T(k)$ over the range
of scales plotted where $T(k)$ changes by 2 to 3 orders of magnitude.  It is
also in wide use and provides a practical reference standard.
By taking the ratio of our results to the BBKS results we can focus on finer
scale features in the spectrum, with the prominent trends due to
matter-radiation equality removed.  

We set the parameter $\Gamma^{*}$ in the BBKS fitting function using
the prescription of Sugiyama \shortcite{Sug}:
\begin{equation}
  \Gamma^{*}\equiv \Omega_0 h\, \exp[-\Ob(1+\sqrt{2h}/\Omega_0)].
\label{eqn:obsug}
\end{equation}
We have labeled this parameter $\Gamma^{*}$ rather than $\Gamma$
to emphasize that it governs the shape of the {\it transfer function\/} alone,
and not the shape of $\Delta^2(k)$ if the initial spectrum is not
scale-invariant.
A more involved expression for $\Gamma^{*}$ is given by
Hu \& Sugiyama~\shortcite{HuSug} (Eqs.~D29, E12), but we shall use the
simple scaling; it has received widespread use elsewhere, and it is
informative to examine its accuracy.
The BBKS fitting function itself is accurate only to about 5\%, even in the
limit $\Ob\to0$, and this accounts for some of the discrepancy shown in
Fig.~\ref{fig:cmblss}.
Finally we should emphasize that Fig.~\ref{fig:cmblss} does not show the
full effect of including baryons in the calculation of the power spectrum.
Compared to a zero baryon model with the same $\Omega_0$, a model with baryons
has less power on all scales smaller than $k\sim 10^{-2}\,h\,{\rm Mpc}^{-1}$.
This suppression of power on small scales is partly accounted for by the
rescaling of $\Gamma$ in Eq.~\ref{eqn:obsug}.
We reiterate that we have divided by the BBKS form purely in order to
remove the trend and focus on the finer features in the power spectrum.

\begin{figure}
\begin{center}
\leavevmode \epsfxsize=3.3in \epsfbox{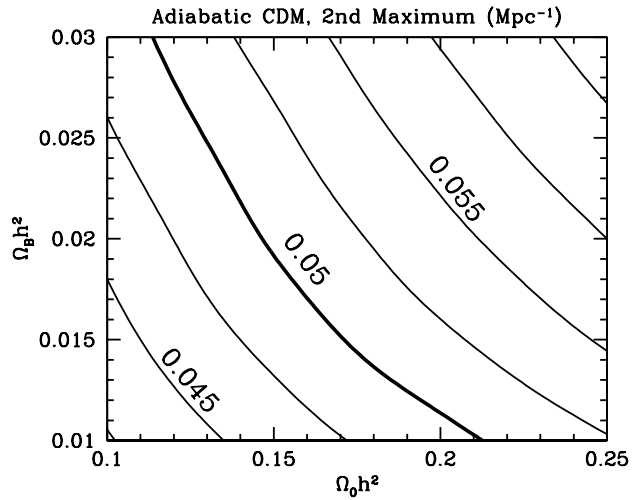}
\end{center}
\caption{The position (in $k$ space in Mpc${}^{-1}$) of the second maximum
in $T(k)$ compared to BBKS, as a function of $\Omega_0h^2$ and $\Ob h^2$.
Here $\Omega_0=\Omega_{\rm CDM}+\Ob$.
For measurements of distances in $\hompc$ the values of the
contours need to be divided by $h$.
Apart from this scaling, the position of the peak is almost independent of
cosmological parameters.}
\label{fig:peakpos}
\end{figure}

Fig.~\ref{fig:cmblss} shows that on scales $k\sim0.1\hompc$,
there is a suppression of power relative to the BBKS fitting function.
Partly this is due to inaccuracies in the BBKS fit, but there is also
a physical effect at work: the baryons are supported by photon pressure
at early times and their fluctuations cannot grow in amplitude until they are
released from the photons.  On small scales the peaks in the matter
and radiation power spectra are out of phase due to velocity overshoot
\cite{SunZel,PreVis,HuSug}.  The growing (decaying) mode of the
perturbations projects primarily onto the velocity (density) of the
perturbation at high $k$.  At larger scales the phases of the peaks
become comparable, as the growing mode is sourced more by the density.
Further discussion is given in Appendix \ref{sec:theory}.  A thorough
discussion of the physics in linear theory and an analytic fitting
function for $T(k)$ are presented by Eisenstein \&
Hu~\shortcite{EisHu} and the application to 100-Mpc clustering by
Eisenstein et al.~\shortcite{EHSS}.  Where our results overlap, they
are in good agreement.

\begin{figure*}
\begin{center}
\leavevmode
\epsfxsize=3.3in \epsfbox{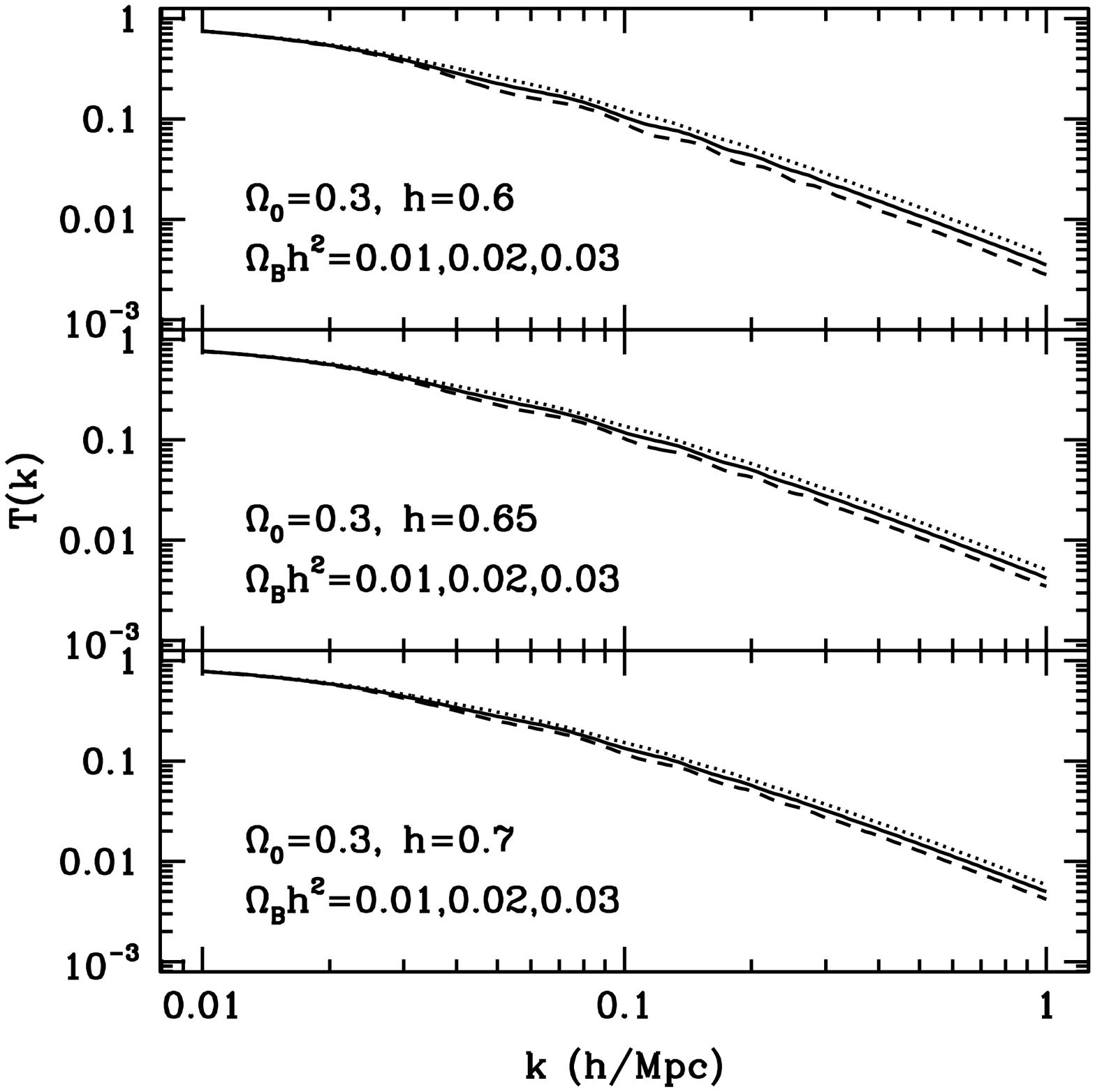}
\epsfxsize=3.3in \epsfbox{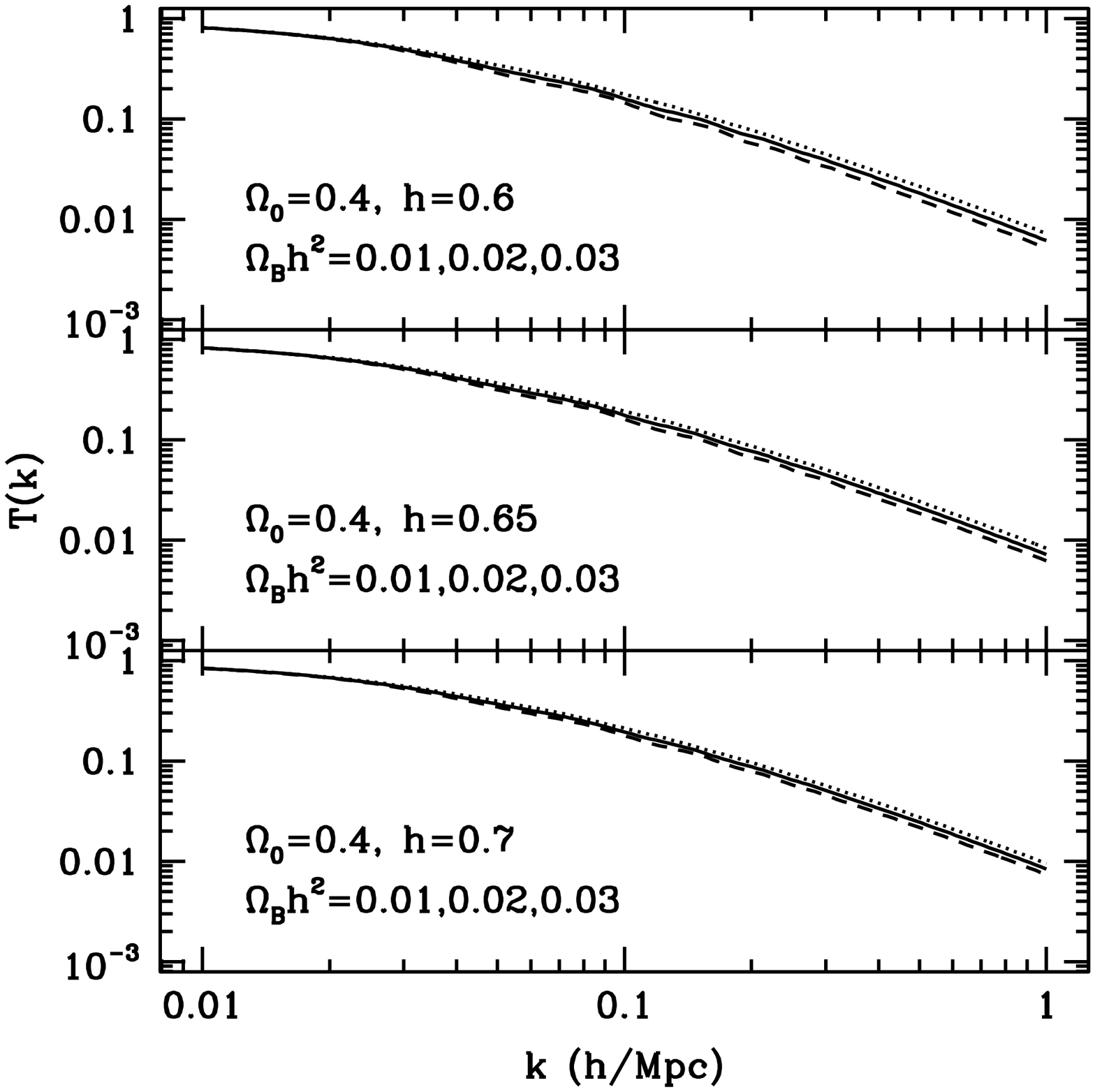}
\end{center}
\caption{The transfer function $T(k)$ for $\Omega_0=0.3$, 0.4 and $h=0.6$,
0.65 and 0.7.  Note that the features in $T(k)$ are quite small on the
scale of the variation of $T(k)$, but prominent when the trend is removed,
as shown in Fig.~\protect\ref{fig:del2}.}
\label{fig:tk}
\end{figure*}

\begin{figure*}
\begin{center}
\leavevmode
\epsfxsize=3.3in \epsfbox{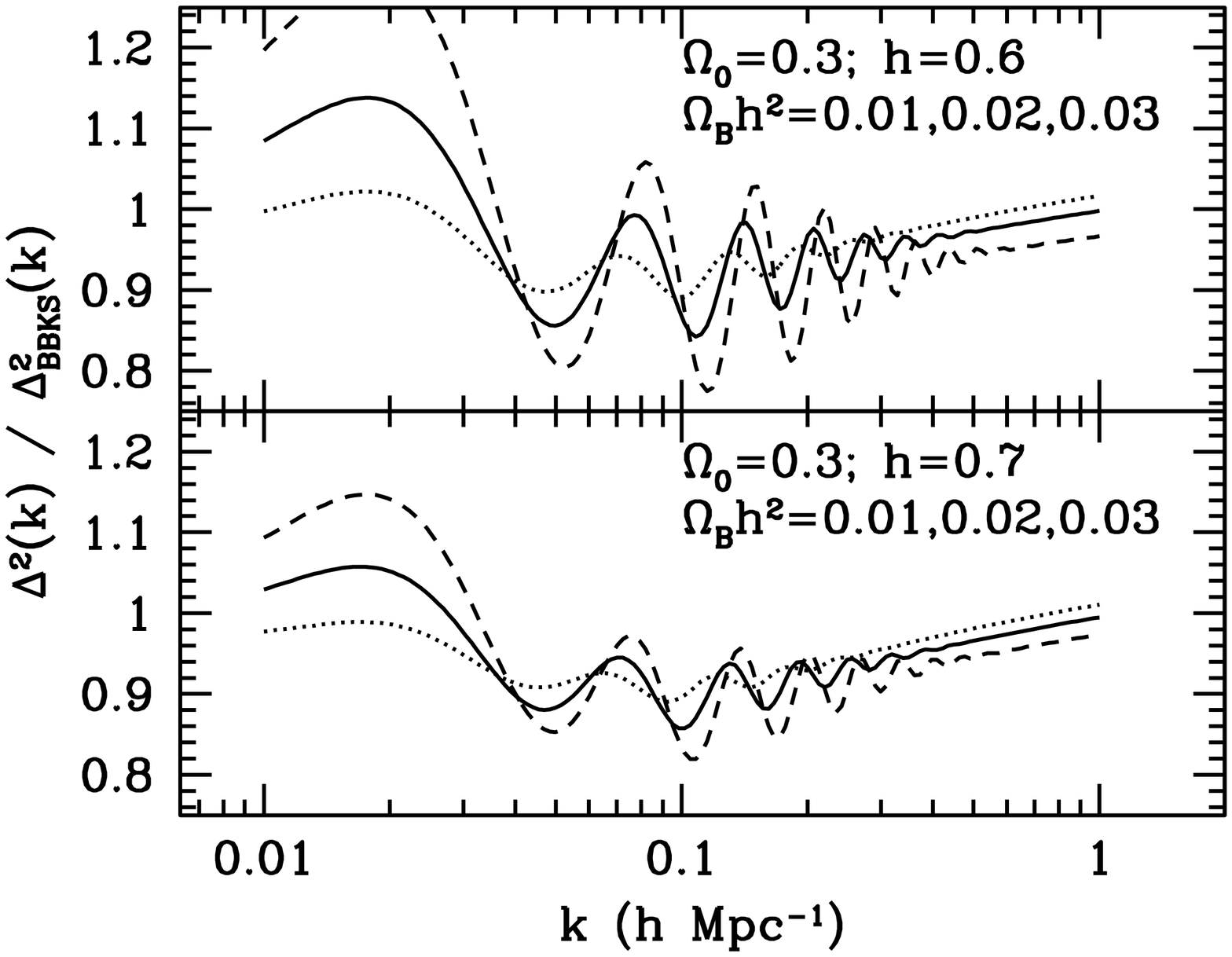}
\epsfxsize=3.3in \epsfbox{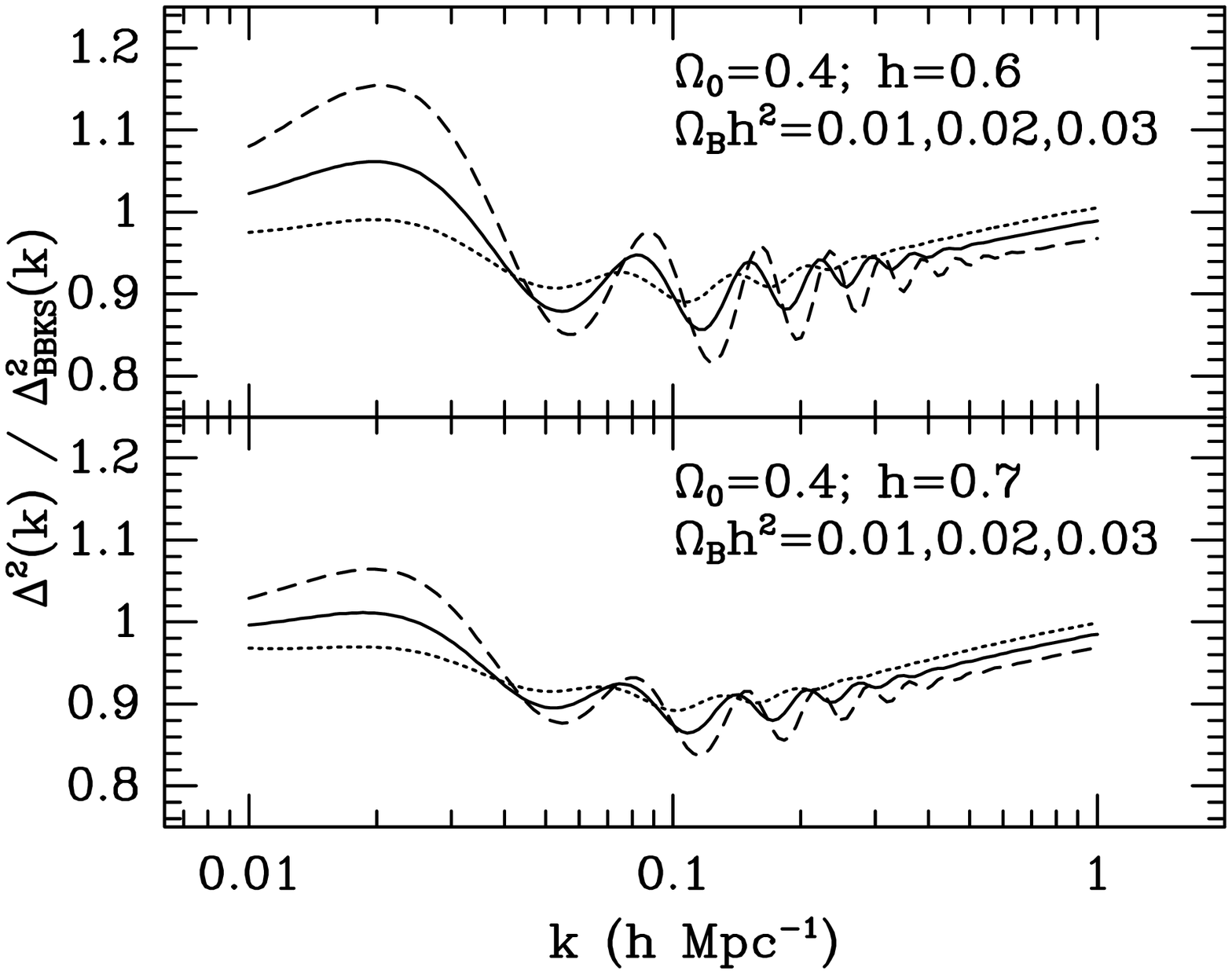}
\end{center}
\caption{The linear theory matter power spectrum  with the trend removed
by dividing by the best fitting BBKS analytic fit to $T(k)$.
Higher $\Ob$ gives larger oscillations.}
\label{fig:del2}
\end{figure*}

The positions of the peaks in $P(k)$ are very insensitive to the cosmological
parameters over the range of interest.
In adiabatic CDM models the second peak is the most likely to be measured
first, since it occurs at a more accessible scale than the first peak.
We show the position (in Mpc${}^{-1}$) of the second peak in adiabatic CDM
models, as a function of $\Omega_0 h^2$ and $\Ob h^2$ in
Fig.~\ref{fig:peakpos}.  The transfer functions for a grid of 35 models in
total were calculated by numerical integration of the coupled Einstein, fluid
and Boltzmann equations from before equality through to the present.  
In Fig.~\ref{fig:tk} we show $T(k)$ for a set of these models. On the absolute
scale of the power spectrum, the features are small. Relative to the
BBKS transfer function, however, the features are prominent.
The peaks in Fig.~\ref{fig:peakpos} were found as the maxima and minima of
$T(k)/T_{\rm BBKS}(k)$ with an additional trend removed to account for the dip
near $k\sim 0.1 \hompc$, as is visible in Fig.~\ref{fig:del2}.
Notice that the position varies by only 10\% for reasonable variations in
these parameters, and can thus be very reliably predicted by models.
For isocurvature models the peaks are shifted by a factor of approximately 1.5
to higher $k$.
The other peaks form an almost harmonic series, as shown in
Fig.~\ref{fig:del2} and discussed in Appendix~\ref{sec:theory}.

\section{Beyond Linear Theory} \label{sec:nonlinear}

The power spectra shown in Figs.~\ref{fig:cmblss} and \ref{fig:del2} 
are computed using linear perturbation theory, as discussed in
Appendix \ref{sec:theory}.
However for {\sl COBE\/} normalized models, even the first few features are
in the trans-linear regime ($\Delta_{\rm L}^2\sim0.1$--1), so it is far from
clear that linear theory will apply to these oscillations. To compute the full
nonlinear power spectrum $\Delta^2$, we have gone to second order in
perturbation theory and performed N-body simulations as described below.
Our conclusion is that the general trend of the non-linear effects is to
suppress the peaks in the non-linear regime
(Figs.~\ref{fig:nonlinearEH}, \ref{fig:nonlinearC},
\ref{fig:nonlinearL} and \ref{fig:nonlinearO}).

\begin{figure}
\begin{center}
\leavevmode \epsfxsize=3.3in \epsfbox{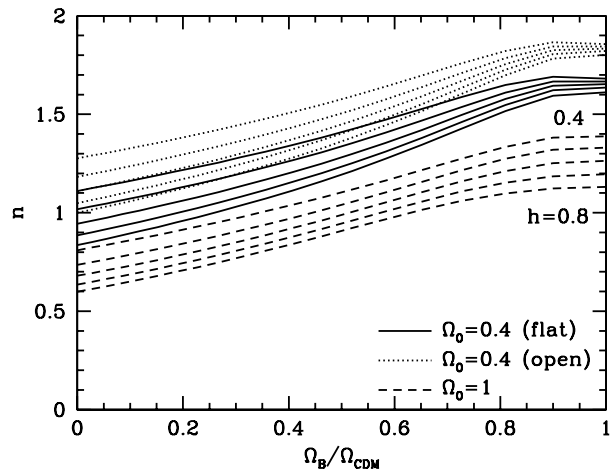}
\end{center}
\caption{The values of primordial spectral index required in order for
CDM models to satisfy the constraints of correct normalization on both
{\sl COBE\/} and cluster scales, for the indicated values of $h$.
This figure assumes no tensor contribution
to the CMB anisotropy.}
\label{fig:tilt_omb}
\end{figure}

\subsection{Normalization}

Since the persistence of the peaks depends on whether they lie above or
below the scale of non-linearity, the absolute normalization of the models
is important.  The {\sl COBE\/} normalization distinguishes between open
and vacuum--dominated models. In the absence of tensor contributions, the
horizon crossing amplitude is
\begin{equation}
\begin{array}{lcl}
\delta_{\rm H} &=& 1.95\times10^{-5}\;
\Omega_0^{-0.35-0.19\ln \Omega_0-0.17 (n-1)}\;\; \times \\
& &\ \ \exp[-(n-1)-0.14 (n-1)^2]
\end{array}
\end{equation}
for open models and
\begin{equation}
\begin{array}{lcl}
\delta_{\rm H} &=&
  1.94\times10^{-5}\; \Omega_0^{-0.785-0.05\ln \Omega_0}\;\;\times\\
  & &\ \ \exp[-0.95(n-1)-0.169 (n-1)^2]
\end{array}
\end{equation}
for flat models \cite{BunWhi}.
Thus if the models are {\sl COBE\/} normalized we would expect the open
models to show more pronounced oscillations.
However one can also normalize the models on small scales
($k\simeq0.2 \hompc$) using the abundance of rich clusters
(see e.g. Pen 1998 and references therein).
In this case the open and $\Lambda$ models have very similar normalizations
\cite{Eke96,VL96,Eke98,VL98}.
We adopt the following simple fitting formulae, which is near to the mean
opinion on this issue:
\begin{equation}
\sigma_8=0.55\, \Omega_0^{-0.56}
\end{equation}
\cite{WEF}.
In what follows, we will generally force the models to fit both {\sl COBE\/}
and the abundance of rich clusters by adjusting the primordial spectral index
$n$ appropriately.
Fig.~\ref{fig:tilt_omb} shows the values of $n$ required in order for CDM
models to satisfy these constraints for various values of $\Omega_0$, $\Ob$
and $h$.
This was calculated using the approximate transfer function of
Eisenstein \& Hu \shortcite{EisHu}, and neglecting any tensor contribution to
CMB anisotropies.
It is apparent that, for reasonable baryon content, $\Omega_0=1$ models
generally require a substantial tilt towards $n<1$, whereas low-$\Omega_0$
models need $n>1$ \cite{WhiSil}.
The smallest degree of tilt is required in the case of $\Lambda$CDM and a
low baryon fraction.

\begin{table}
\begin{center}
\begin{tabular}{l|cccccc}
Model & $\Omega_0$ & $\Ob h^2$ & $h$ & n & $\Gamma^{*}$
& $\sigma_8$ \\ \hline
tCDM          & 1.0 & 0.0375 & 0.50 & 0.748 & 0.370 & 0.55 \\
OCDM          & 0.4 & 0.0300 & 0.65 & 1.258 & 0.198 & 0.92 \\
$\Lambda$CDM  & 0.4 & 0.0300 & 0.65 & 1.030 & 0.198 & 0.92 \\
EHSS          & 1.0 & 0.1440 & 0.60 & 0.889 & 0.259 & 0.55
\end{tabular}
\end{center}
\caption{Parameters for the cosmological models considered in the text.
The column labeled $\Gamma^{*}$ is the parameter $\Gamma$ of the BBKS
transfer function which best fits the numerical transfer function of
the model, as determined by Sugiyama \protect\shortcite{Sug}.
These are not the values of $\Gamma$ which would be inferred from $\Delta^2(k)$
since $n\ne1$.
The model EHSS is the high-$\Omega_{\rm B}$ model suggested by
Eisenstein et al.~\protect\shortcite{EHSS}, with $n$ reduced in
order to fit the cluster abundance.}
\label{tab:models}
\end{table}

\subsection{Perturbation Theory}

\begin{figure}
\begin{center}
\leavevmode \epsfxsize=3.3in \epsfbox{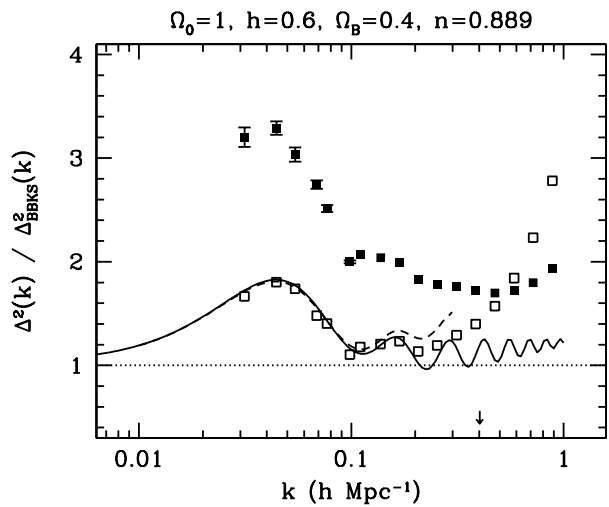}
\end{center}
\caption{The effect of non-linear evolution on the oscillations in the power
spectrum.  We show (solid line) the ratio of the linear power spectrum for
the model suggested by Eisenstein et al.~\protect\shortcite{EHSS} with
$\Omega_0=1$, $h=0.6$, $\Ob=0.4$ and $n=0.889$ (we have lowered $n$ to
provide a better fit to the cluster abundance) to a BBKS model.
The BBKS transfer function has $\Gamma^{*}=0.259$ and the model is
{\sl COBE\/} normalized.
Note that the BBKS transfer function with $\Gamma^{*}$ as provided by
Sugiyama (1995) does not provide a good fit to the numerical $T(k)$.  At
small scales a much better fit is provided by $\Gamma^{*}=0.265$.
The dashed line shows the ratio including 2nd order corrections, the small
arrow near $k\simeq0.4 \hompc$ indicates where $\Delta^2(k)=1$.
The open squares are the results from N-body simulations with the CDM
spectrum in real space, solid squares the results in redshift space.
Note that the oscillations for $k\protect\ga 0.1\hompc$ are washed out.}
\label{fig:nonlinearEH}
\end{figure}

Since the features we are interested in occur in the regime where
$\Delta^2(k)\sim1$ we first computed the correction to the linear theory
result using second order perturbation theory
\cite{Jusz81,Vish83,Mak92,Jain94}.
The basic method involves evaluating two convolution integrals numerically,
\begin{eqnarray}
&&\Delta^2_{(2)}(k) = \Delta_{\rm L}^2(k) \nonumber \\
&&\ \ +\int_0^\infty d\ln q\; \Delta_{\rm L}^2(q)\, \int_{-1}^1 d\mu\; 
\Delta_{\rm L}^2(\vert{\bf k}-{\bf q}\vert)\, K_1({\bf k}, {\bf q})
\nonumber \\
&&\ \ +\Delta_{\rm L}^2(k)\, \int_0^\infty d\ln q\; \Delta_{\rm L}^2(q)
\, K_2(k, q),
\end{eqnarray}
where $\Delta^2_{(2)}(k)$ is the power
spectrum including second order corrections, and $K_1({\bf k}, {\bf q})$ and
$K_2(k, q)$ are integration kernels. These are given by
\begin{eqnarray}
K_1({\bf k}, {\bf q})&=&\frac{1}{196}\frac{k^3}{q^4
\vert{\bf k}-{\bf q}\vert^7}\;\;\times \nonumber \\
&&(3q^2k^2+7qk^3\mu -10q^2k^2\mu^2)^2,
\end{eqnarray}
where $\mu={\bf k}\cdot{\bf q}/ (kq)$, and
\begin{eqnarray}
K_2(k, q)&=&\frac{1}{252}\frac{k^2}{q^2}
\Bigl[6\frac{k^2}{q^2}-79+50\frac{q^2}{k^2}-21\frac{q^4}{k^4}
\nonumber \\
&&+\frac{3}{2k^5q^3}(q^2-k^2)^3(7q^2+2k^2)
\nonumber \\
&&\times\ln\left(\frac{k+q}{\vert k-q \vert}\right)\Bigr].
\end{eqnarray}
The perturbation expansion formally breaks down for large $\Delta_{\rm L}^2(k)$.
In principle the integration over $q$ should be terminated at some maximum
wavenumber; in practice, the results are not very sensitive to the upper
limit. We note that, for an Einstein--deSitter universe, the second
order correction to the power spectrum evolves as $a^4$, while the power
spectrum in linear order evolves as $a^2$, where $a$ is the expansion factor.
We adopt the same expressions for the open models below, since the
additional corrections to the second order terms appear small \cite{Cat95}.

One may also compare the second--order calculation with the fully nonlinear
scaling formula developed by Peacock \& Dodds~\shortcite{PD96}.
Extension of this formalism to treat oscillatory spectra is difficult
since there is no unique definition of a local spectral index.
A naive application of the results, using the spectral index from the
best-fitting BBKS spectrum and the full oscillatory power spectrum in
$\Delta^2_{\rm L}$, results in a very poor match to the exact non-linear
spectrum.  We have not pursued this matter further.

\subsection{N-body models}

\begin{figure}
\begin{center}
\leavevmode \epsfxsize=3.3in \epsfbox{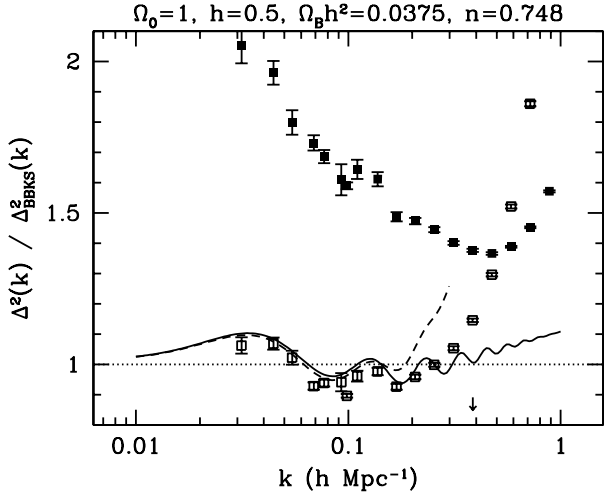}
\end{center}
\caption{As in Fig.~\protect\ref{fig:nonlinearEH} but for a model with
$\Omega_0=1$, $h=0.50$, $\Ob=0.15$ and $n=0.748$.
The BBKS transfer function has $\Gamma^{*}=0.37$ and the model is
{\sl COBE\/} normalized.}
\label{fig:nonlinearC}
\end{figure}

\begin{figure}
\begin{center}
\leavevmode \epsfxsize=3.3in \epsfbox{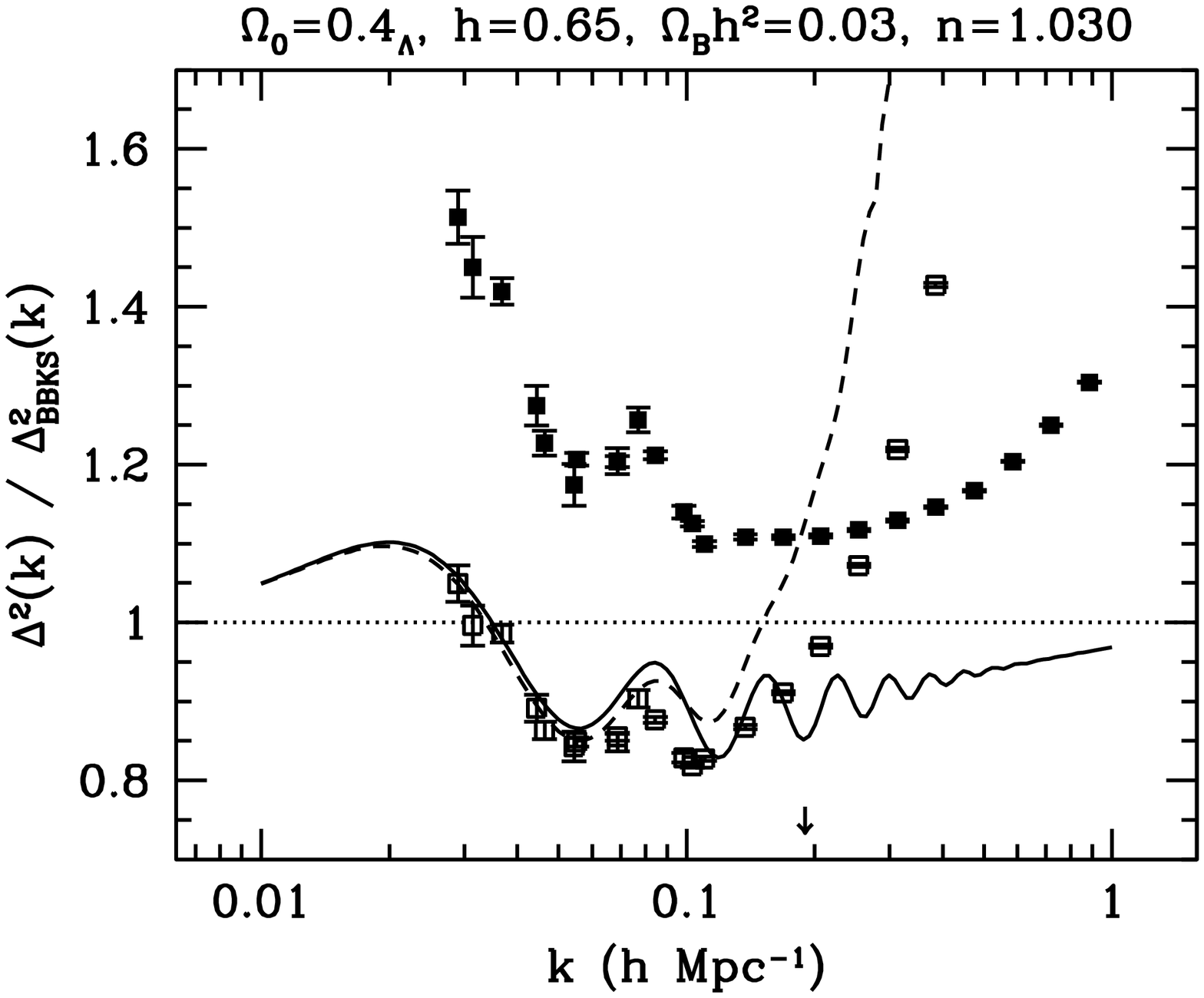}
\end{center}
\caption{As in Fig.~\protect\ref{fig:nonlinearEH} but for a model with
$\Omega_0=0.4$, $\Omega_\Lambda=0.6$, $h=0.65$, $\Ob h^2=0.03$ and $n=1.030$.
The BBKS transfer function has $\Gamma^{*}=0.198$ and the model is
{\sl COBE\/} normalized.}
\label{fig:nonlinearL}
\end{figure}

\begin{figure}
\begin{center}
\leavevmode \epsfxsize=3.3in \epsfbox{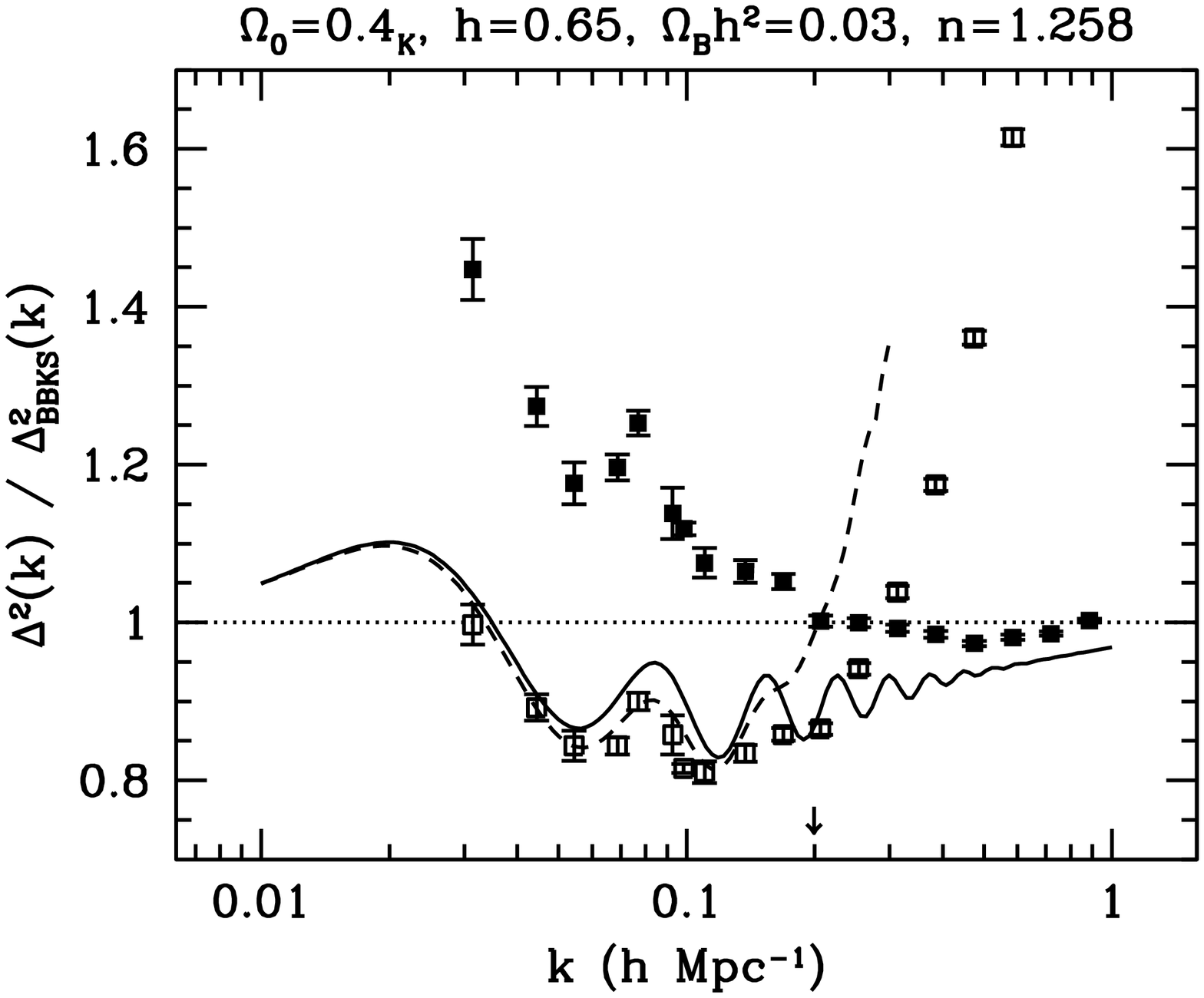}
\end{center}
\caption{As in Fig.~\protect\ref{fig:nonlinearEH} but for a model with
$\Omega_0=0.4$, $\Omega_\Lambda=0$, $h=0.65$, $\Ob h^2=0.03$ and $n=1.258$.
The BBKS transfer function has $\Gamma^{*}=0.198$ and the model is
{\sl COBE\/} normalized.}
\label{fig:nonlinearO}
\end{figure}

In order to find the exact behaviour of the spectrum, we ran a series of
N-body simulations using two P$^3$M codes.  The first was an adaptation of
Hugh Couchman's AP$^3$M code 
(Couchman~1991; see also Peacock \& Dodds~1996)
and the second was the Hydra N-body, hydrodynamics code, run in collisionless
mode \cite{Hydra}.  The agreement in $\Delta^2(k)$ between both of these codes
and an independent PM code (see Appendix \ref{sec:pmcode}) was very good,
indicating that the numerical results were stable.
Note that these codes treat all matter as cold, whether or not some fraction
is baryonic or HDM.
Such differences have a strong effect on small scales,
but for $k\la 1\hompc$ are negligible.

Having determined that the PM code adequately calculated the non-linear
power spectrum on the scales of interest, we used its superior speed to
allow the averaging over many realizations.
For each model the PM code was run many times (typically $10^3$ in $64^3$ mode
for the low-$k$ modes and few$\times 10^2$ in $128^3$ mode for the high-$k$
modes) with a starting redshift of $1+z=20$, and different Gaussian initial
conditions.
The resulting determinations of $\Delta^2(k)$ were averaged over the runs.
An alternative procedure designed to reduce scatter at low-$k$, allowing
random phases but with fixed amplitudes \cite{PD96}, was found not to be
sufficiently accurate for our purposes.
With a careful treatment of binning effects, we believe our results for the
power spectrum are accurate to 1\%.

\subsection{Real--space power spectra}

The results of these procedures are shown in
Figs.~\ref{fig:nonlinearEH}--\ref{fig:nonlinearO}.
The high-$k$ oscillations in $\Delta^2(k)$ are suppressed once second order
effects are included. There is an additional reduction of power on somewhat
longer scales ($k\la 0.1\hompc$). Second order perturbation theory provides
a reasonable, but not highly accurate, description of the non-linear effects.
It successfully predicts the scale of complete suppression of the peaks, but
breaks down at $\Delta^2(k)\la1$.

The clear message of these figures is that only one unambiguous
baryonic feature in the spectrum is expected. The first baryonic
peak relative to BBKS is hardly altered by nonlinear evolution,
but it is not clear that it is detectable. As we discuss below,
the likely future errors on the power spectrum at these very
large scales will not allow such a 10\% feature to be picked
out. In any case, it is very broad in $k$-space extent, and
will depend rather sensitively on which BBKS model is used as a reference.
In contrast, the second peak is relatively narrow, being confined
to roughly a factor 1.3 in $k$. This feature is usually near
the maximum wavenumber at which it can survive nonlinear damping,
and the observational errors are much smaller. Finally, we have shown
in \S2 that the position of this peak is quite robust once
$h$ is determined. Future efforts to prove the importance of
baryons in determining the form of the fluctuation should
therefore concentrate on this signature.

The results so far have been entirely in real space.
However, the only statistics which are independent of redshift-space effects
are projected statistics, e.g.~the projected correlation function:
\begin{equation}
\Xi(r)=\int_{-\infty}^\infty \xi[(r^2+x^2)^{1/2}]\; dx,
\end{equation}
or the angular correlation function $w(\theta)$.
Integration over a wide selection function damps the visibility of
oscillations. With typical survey selection functions the features are
suppressed to an unmeasurable level.  To avoid this one must use sub-samples
with a narrow distribution in depth.
We will discuss the Fourier space analog of $\Xi(r)$ calculated on narrow
shells selected using photometric redshifts in \S\ref{sec:measure}.

\subsection{Redshift--Space Power--Spectra}

The redshift-space power spectrum is distorted from the real-space power
spectrum by peculiar velocities.  On large scales this leads to an increase
in power \cite{Kaiser} because objects stream towards overdensities.
On short scales virialized motions suppress clustering in redshift space.

We have computed the power spectrum in redshift space, $\Delta_s^2(k)$, from
the N-body codes. For simplicity, an infinitely distant observer is assumed;
i.e., the redshift distortions are along a single Cartesian axis of the box.
The results are in reasonable agreement with analytical models of the redshift
space distortion which involve convolving the Kaiser~\shortcite{Kaiser}
result with an exponential on small scales \cite{ColFisWei}.
However, the transition from enhancement to suppression of $\Delta^2(k)$ starts
to set in near $k\sim0.1-0.2\hompc$, and by $k\sim0.3-0.5\hompc$ the
suppression is severe. Since the position of the features we are proposing
to observe are near $k\sim0.1\hompc$, an interpretation of the power spectrum
in redshift space requires a careful modeling of the redshift-space
distortions.

We show $\Delta_s^2(k)$ as the solid squares in
Figs.~\ref{fig:nonlinearEH}, \ref{fig:nonlinearC}, \ref{fig:nonlinearL}
and \ref{fig:nonlinearO}.
Note that the appearance of the baryonic features are altered by the redshift
space distortions, but persist in a recognizable form.
As in the real-space results, the only feature that will be robust with
respect to how the smooth underlying spectrum is defined will be the second
peak, which is expected to give a roughly 10\% boost to the power in a range
of $k$ of about a factor 1.3 centred on $k\simeq 0.055\, \rm Mpc^{-1}$.
The detection of this signature will be an observational challenge but,
as we discuss below, it may be feasible.

\section{Fitting clustering data} \label{sec:clustering}

So far our discussion has been confined to a firm prediction for the CDM
family of models:
for large $\Ob$ and low $\Omega_0$, we expect to see oscillations
in the matter power spectrum arising from the same mechanism (acoustic waves
before decoupling) which produced the peaks in the CMB spectrum, for which
evidence is mounting.
We next consider the possibility that tentative evidence for the oscillations
has been found in recent surveys.

There are several surveys which have quoted evidence for excess power
on 100-Mpc scales. The first was the Broadhurst et al.~\shortcite{BEKS}
pencil beam surveys, which quoted an excess of power on $128\mpcoh$ wavelengths.
The 2D analysis of the LCRS survey \cite{Lanetal} also reported
a statistically significant bump in the power spectrum at
$k\sim0.06 \hompc$ (but see Bromley \& Press 1998).
Recently Einasto et al.~\shortcite{EinA,EinB} have claimed to
find $120\mpcoh$ periodicity in the 3D distribution of superclusters.
Evidence for a distinct kind of excess power on large scales is the broad bump
detected around $k\simeq 0.1\hompc$ in the APM and IRAS
real--space data \cite{Saun92,Mad96,Pea}.

The intriguing possibility is that these surveys may be indicating the
existence of acoustic oscillations in the underlying matter power
spectrum.  From our earlier discussion it appears that one cannot explain
these features from the linear-theory mass power spectrum of an
adiabatic CDM model with $\Ob$ constrained by standard BBN.
What would be required to explain such features?
This question has been addressed recently in linear theory by
Eisenstein et al.~\shortcite{EHSS}.
They identify two models which might fit the data, one with low $\Omega_0$
and one with high $\Omega_0$.
As expected from the discussion in \S\ref{sec:linear},
the parameters required for either of these models do not appear credible
{}from other considerations (e.g.,~BBN constraints on the baryon fraction).
However these models provide, by construction, features in the places for
which there is (controversial) evidence as discussed above.

To see how the features are affected by non-linearities, we have run N-body
simulations of the high-$\Omega_0$ model
($\Omega_0=1$, $\Ob=0.4$, $h=0.6$, $n=0.889$).
The result is shown in Fig.~\ref{fig:nonlinearEH}, and differs somewhat from
the other models we have investigated.
Because of the extreme parameter choices in this case, the baryonic features
are at somewhat larger $k$.  This aspect allows the model to place a broad
feature of amplitude roughly a factor 1.8 at the desired $k\simeq 0.05\hompc$
-- but the second and subsequent peaks are then at sufficiently high $k$ that
they are destroyed by nonlinearities.
There has to be serious doubt over whether such a relatively broad 3D feature
could cause a significant effect in lower-dimensional surveys \cite{KP91},
but in any case we have shown that smaller-scale harmonics will not survive
in order to give a test of such a hypothesis.

An effect that may alter the relative heights of the peaks is the relative
bias between different galaxy populations that have different redshift
distributions.
By treating the two populations as a single population, a scale dependent
bias in the power spectrum will be introduced in the measured 2D power
spectrum, because the angular scale corresponding to a given 3D clustering
strength will differ for the two populations.
For example, a luminosity--dependent clustering amplitude could lead to
such an effect.

The standard prescription for calculating the power spectrum from a
survey implicitly assumes that the clustering is independent of the
luminosity. There is evidence \cite{dms,LCRS,BPLK,WDP} that this is not
the case. We have tested the effects of two populations with
differing bias factors and selection functions on the calculation of
the power spectrum, presuming the two populations are treated as a
single population. We find that in all cases the effect is weak. The
only case in which a large effect can be seen is if one of the
populations has a tight selection function in $z$, though we regard
this case as artificial. For realistic selection functions the
scale dependent bias introduced by two populations with $b_1/b_2\la 2$
is at the 1\% level. We give details in Appendix \ref{sec:sdbias}.

While the distribution of power in a single mode is exponential even in the
presence of strong nonlinearities \cite{FB95}, in a finite-volume survey
nonlinear contributions may increase the dispersion in the power estimate above
the exponential expectation \cite{Amen94}. This may be viewed as a consequence
of the finite number of statistically independent spatial cells in any given
survey. Amendola~\shortcite{Amen94} showed that this may be a particularly
strong effect in pencil-beam surveys. An extension of the analysis to higher
dimensions shows that the non-exponential terms decrease in size inversely with
the number of independent cells within the survey. The effect is also in part
a consequence of the small number of modes available to measure power at low
$k$. Since modes are angle-averaged within frequency bands in multidimensional
surveys, a large number of independent modes will contribute to band-averaged
power spectrum estimates even on large scales.
This ensures that the power estimates will follow a $\chi^2$ distribution
with a large number of degrees of freedom.
In the limit that the number of independent modes $m\rightarrow\infty$,
$[P(k)/\langle P(k)\rangle]^{1/2}$, where $P(k)$ is the band-averaged power
near $k$ and $\langle P(k)\rangle$ is the expected power, tends to a normal
Gaussian deviate with unit mean and variance $1/ 4m$ \cite{KK59}.
Thus it would appear that any features in the power spectrum on the scales
of interest are not likely to be large deviations from the mean power on
those scales.

\section{Future observational Prospects} \label{sec:measure}

Since the features that we are proposing to measure are only small
perturbations to the power spectrum, they are difficult to see in
current surveys (see \S\ref{sec:clustering}).
However both the AAT-2dF and the Sloan Digital Sky Survey (SDSS) will
soon enhance our knowledge of the 3D geometry of the local universe.
We focus here on the SDSS.

\subsection{3D Surveys}

The SDSS Northern Polar Cap redshift survey will contain $10^6$ galaxies with
a median distance of $\sim350 \mpcoh$. An additional $\sim5\times10^7$
galaxies will be measured in 5 bands in a photometric survey $\sim5$ magnitudes
deeper than the redshift catalog. It will be possible to measure the 2D power
spectrum (or angular correlation function) using this deeper survey. It will
additionally be possible to extract measurements of the 3D power spectrum
from the larger catalog by the use of photometric redshifts.

A critical issue for the detection of baryonic features is the expected
precision of these power estimates.  Feldman et al. (1994; FKP) studied this
question in some detail, and we now summarize their main results
(see also Tegmark et al.~1998).
Throughout it is assumed that on the scales of interest the modes remain
approximately independent.  While this is roughly true, in order to attain
the high levels of precision of which the SDSS is capable this assumption may
need to be improved \cite{MeiWhi}.
The FKP formula for the fractional variance in the power measured for a
Gaussian field by averaging modes in some $k$-space volume $V_k$ is their
equation (2.3.2):
\begin{equation}
\left({\sigma_P\over P}\right)^2 =
{(2\pi)^3\over V_k}\;
{ \int d^3r\; {\bar n}^4 w^4[1+1/{\bar n}P]^2 \over
[\int d^3r\; {\bar n}^2 w^2]^2},
\end{equation}
where ${\bar n}({\bf r})$ is the mean density defined by the survey selection,
and $w({\bf r})$ is a weight function, which should be set to
$w=[1+{\bar n} P]^{-1}$ for minimum variance.
This formula assumes that the $k$-space region is large in all directions
compared to the window function resulting from transforming the survey
geometry.
This assumption breaks down at the very largest scales;
see Goldberg \& Strauss~\shortcite{GolStr} for a more sophisticated analysis.
Also, in the case where $V_k$ is a shell in $k$ space, rather than an
isolated region, the above formula should be multiplied by a factor 2 to
allow for the fact that the density field is real (FKP equation 2.4.6).

In the simple case of a large survey where shot noise is negligible,
${\bar n}P \gg 1$, this expression becomes simply
\begin{equation}
\left({\sigma_P\over P}\right)^2= [2] \times (2\pi)^3 (V_k V)^{-1},
\end{equation}
where $V$ is the survey volume.
This reflects the fact that the density of states in $k$ space is
$(2\pi)^3/V$.
The weight function automatically cuts the survey off at the point where
shot noise is starting to become important.
We may choose, however, to analyze a volume-limited subsample for which the
shot noise is negligible.

FKP also give an expression (their 2.2.6) for the covariance of the power
estimates at different wavenumbers.
If shot noise is negligible, this says that the two-point correlation function
for fluctuations in power is just the Fourier transform of $\bar n^2 w^2$.
However, for a shell whose width $\Delta k$ is large enough for the above
expression for the variance in the shell-averaged power to apply, this
correlation function is guaranteed to be small.  The power estimates in
adjacent bins will thus be effectively uncorrelated.

As an example, the FKP effective volume for the Sloan survey is approximately
$0.17\,(h^{-1}\rm Gpc)^3$, taking $P=5000\, (h^{-1}\rm Mpc)^3$ as a typical
reference power, and using the selection function given in
Gunn \& Weinberg~\shortcite{GW}.
This is equivalent to uniformly sampling the volume out to $z\simeq 0.21$;
for an illustrative discussion, we henceforth assume that the Sloan survey
will sample this volume in a way that is effectively independent of shot
noise (we do include the shot-noise component in the error bars, but it is a
very small correction).
For $\Omega=1$ and the Sloan area of $\pi$~sr, the comoving volume is
equivalent to that of a sphere of radius approximately $R=340\mpcoh$.
If the survey window was in fact perfectly isotropic, the power correlation
function would be 
\begin{equation}
  \langle \delta P(k) \delta P(k+\Delta k) \rangle
  \propto |f(R\Delta k)|^2
\end{equation}
where $f(y)\equiv 3(\sin y - y \cos y)/y^3$.
This has its first zero at $\Delta k\simeq 4.5/R$, suggesting that samples
of the power at a spacing of $\ga 0.013\hompc$ will be uncorrelated.
The transform of the real conical window function is in practice about 15
per cent broader than would be calculated just on the basis of the effective
volume of the survey.
Overall, this implies that we can treat estimates of power separated by more
than about $0.015\hompc$ as being effectively completely uncorrelated.
Note that this is a rather more conservative criterion than the
$\Delta k=0.0043\hompc$ assumed by Goldberg \& Strauss~\shortcite{GolStr}.
According to our calculations, the power correlation coefficient at this
separation should be around 0.5.
This may explain why they found slightly too small uncertainties in model
fits to the power spectra when using FKP error bars.

We show in Fig.~\ref{fig:fitbar} the error bars which the SDSS Northern Polar
Cap (NPC) redshift survey should be able to produce on the power spectrum.
The error bars are for bins with a constant spacing of $\Delta k=0.015\hompc$
in wavenumber.
The above formula suggests a fractional power error of
\begin{equation}
 {\sigma_P\over P}=\sqrt{4\pi^2\over V k^2\Delta k} = (k/0.0039\hompc)^{-1}.
\end{equation}

The NPC survey may be extended in several ways which will enhance its ability
to measure $\Delta^2(k)$ at low $k$.
First will be by autocorrelation including the Southern Redshift Survey, which
provides several long baselines between the north and south to constrain the
large-scale power.
Secondly a deeper survey will be done using Bright Red Galaxies, which could
reach to an effective depth of close to 1 Gpc. This will at least double the
number of independent modes compared with the NPC survey, with a corresponding
reduction of the errors by a factor of 2. In Fig.~\ref{fig:fitbar} we show
representative error bars corresponding to $\Delta k=0.0075\hompc$ and
$\sigma_P/P=(k/0.00195\hompc)^{-1}$.

Because of the great depth to which QSO sources may be detected, it might be
hoped that they could be used to obtain an even more precise measure of the
power spectrum. Unfortunately their sparse numbers would restrict any such
measurement to being shot noise dominated in the forseable future. For
example, Warren, Hewett, \& Osmer (1994) find the proper density of QSOs
in a sample slightly deeper ($m_{or}<20$) than either the SDSS or 2dF samples
peaks at a value of $\bar n_{QSO}\approx10^{-3.5} h^3 {\rm Mpc}^{-3}$ at
$z=3.3$ ($q_0=0.5$). A fiducial value for the power spectrum today of
$P=5000 h^{-3} {\rm Mpc}^3$ corresponds to $P\approx300 h^{-3} {\rm Mpc}^3$ at
$z=3.3$, an order of magnitude smaller than the shot noise contribution
$\bar n_{QSO}^{-1}\approx3300$. It is quite likely that the QSOs are a biased
tracer of the underlying mass fluctuations. Still, the number of QSOs must be
at least a factor $100/ b_{QSO}^2$ larger, where $b_{QSO}$ is the bias
factor, to detect the baryonic signature in the power spectrum. At this time
it is unclear whether QSOs even exist in such numbers for plausible bias
factors.

Finally one could imagine using photometric redshifts for the full
$5\times10^7$ galaxies in the photometric catalogue to compute $\Delta^2(k)$
at low $k$ where the effects of the redshift errors are less important.
The larger number of galaxies allows one to probe deeper, and reduces the
errors at low $k$ as in the BRG sample.
How well this may be done in practice will depend on the redshift distribution
of the galaxies.  The distance errors from photometric redshifts at low $z$
are on the order of $\sigma\sim 50\,h^{-1}\,{\rm Mpc}$ \cite{BCSB}.
The power spectrum is strongly affected when $k\sigma\ge 1$,
or $k\ge 0.02\,h\,{\rm Mpc}^{-1}$.
On the other hand, if there is a sufficient number of galaxies to high
redshifts, say $z\approx1$, then the decreased comoving differential
path length with redshift will increase the usable region in (comoving)
$k$-space up to $k\approx0.04\hompc$ if errors as small as $\delta z=0.02$ may
be achieved to these high redshifts \cite{Con95}.
This may be adequate for a detection of the primary, and largest, peak.
Systematic errors are still potentially large in this range, but they may
be sufficiently controllable.
How to deal with the effect of evolution in these very deep surveys, however,
needs further study.

These questions aside, the error bars shown in Fig.~\ref{fig:fitbar} represent
a conservative estimate which may be able to be improved upon with more work.
Clearly the ability of the SDSS to measure the oscillations in $\Delta^2(k)$
will depend on the underlying cosmological parameters.

\subsection{Degenerate models}

\begin{figure}
\begin{center}
\leavevmode
\epsfxsize=3.3in \epsfbox{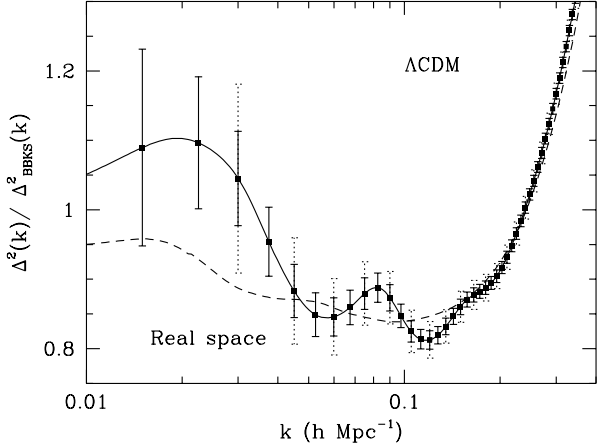}
\end{center}
\caption{A comparison of the fiducial $\Lambda$CDM model (solid line)
with a variant of the model having no baryons (dashed line).
The dotted error bars are estimates of $1\sigma$ errors in uncorrelated
bins which should be measurable with the SDSS Northern Polar Cap survey
assuming the underlying model is $\Lambda$CDM.  The smaller solid error
bars are for the Bright Red Galaxy sample.  The BRG sample will be able
to detect the baryon signature in the quasi-linear to linear regime only
if the errors shown for $k<0.045\, h {\rm Mpc}^{-1}$ are realized.
The fiducial model has $\Omega_0=0.4=1-\Omega_\Lambda$,
$\Omega_{\rm B}h^2=0.03$, $h=0.65$, and $n=1.030$. The non-baryonic model
has $n=0.96$, $h=0.56$ and $\Omega_{\rm B}=0$.
Both models are COBE normalized, and divided by the same BBKS model fit
used in Fig.~\ref{fig:nonlinearL}. (The small ``oscillations'' in the
non-baryonic model at $0.03<k<0.06\hompc$ are due to the limited accuracy of
the non-linear calculation at low $k$, and are not physical.)}
\label{fig:fitbar}
\end{figure}

We quantify the ability of the SDSS to detect the presence of the acoustic
oscillations by computing the expected $\chi^2$ between a power spectrum with
oscillations and a corresponding limiting case with no baryons.
In so doing, we assume that non-linear mode coupling can be neglected and the
power estimates measured at the various sampling points are statistically
independent of each other.
Statistical independence is a stronger assumption than the result proved
above, which is that sufficiently well-separated power estimates are
uncorrelated.
In practice, the difference appears to be small, and the reduced
$\chi^2$ values for the fit to the exact spectrum are very close to
unity. We therefore neglect this technical distinction.
Our neglect of non-linear mode coupling is slightly optimistic, but will
not alter our results too much provided we work at $k<0.1\,h\,{\rm Mpc}^{-1}$
\cite{MeiWhi}.

In Fig.~\ref{fig:fitbar}, we show a comparison between the fiducial
$\Lambda$CDM model and a variant with $n=0.96$, $h=0.56$ and
$\Omega_{\rm B}=0$. The non-linear spectrum was again calculated from the
N-body code, even though there are no oscillations to render the
Peacock \& Dodds \shortcite{PD96} procedure inapplicable. We found that the
non-linear scaling prescription was not always accurate enough for
our purposes, which may be an important consideration when analyzing future
high precision surveys.
The model, which we fit by eye, has the large-scale clustering amplitude
held fixed at the COBE central value \cite{BunWhi}.  We allowed the
tilt and ``shape'' to vary to obtain the closest fit to our $\Lambda$CDM model.
The data points and errors are those expected for the SDSS Bright Red Galaxy
sample and the North Polar Cap sample, assuming $\Lambda$CDM is the true
underlying model. The NPC sample is able to distinguish the two models at
the $3\sigma$ level only for $k>0.33\, h {\rm Mpc}^{-1}$.
The greater depth of the BRG sample permits a clear distinction at the
$3\sigma$ level to be made between the two models from the
$k<0.045\, h {\rm Mpc}^{-1}$ measurements.
These measurements, however, may be difficult to achieve because of
correlations in the photometry on these scales either due to intrinsic
measurement error or as a result of residual correlations in Galactic
extinction. Confining the comparison to larger $k$ values requires
points with $k>0.13\, h {\rm Mpc}^{-1}$, for which $\Delta^2(k)>0.5$, for the
BRG sample to separate the models at the $3\sigma$ level. Because the effects
of bias and redshift-space distortions
become difficult to estimate on these scales, obtaining the low $k$
measurements may be required.
Alternatively, under the assumption that the dominant errors in both
experiments are statistical, it is possible that a combined fit of the
CMB anisotropy data from MAP and the SDSS data will break any degeneracy
in the quasi-linear regime. This has been shown using a linear analysis by
Eisenstein, Hu \& Tegmark (1998).
Note, however, that the second harmonic in Figure~\ref{fig:fitbar} is
invisible even to the BRG sample:\ over $0.045<k<0.1\hompc$ the BRG sample
rejects the non-baryonic model with a confidence of only 90\%. If the depth
of the BRG is able to reach fully to 1 Gpc, then the situation improves. With
the increased number of modes and precision, the BRG sample would then be
able to distinguish the baryonic model from the non-baryonic model over the
range $0.045<k<0.1\hompc$ with a confidence of 99.93\%. This
optimistically presumes that evolution of BRGs over such a broad range in
ages is well understood. It also neglects intrinsic correlations in the
power spectrum, which are moderately large even near $k=0.1\hompc$
\cite{MeiWhi}.

\begin{figure}
\begin{center}
\leavevmode
\epsfxsize=3.3in \epsfbox{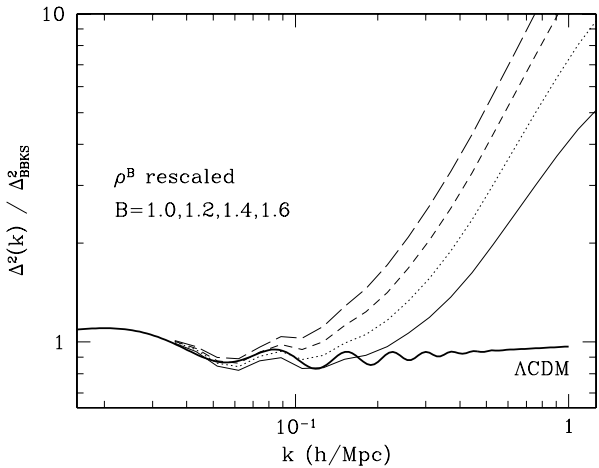}
\end{center}
\caption{The power spectrum of the filtered density field $\rho\to\rho^B$
for $B=1.0, 1.2, 1.4$ and 1.6 in the $\Lambda$CDM model.  The $y$-axis is
the ratio of $\Delta^2(k)$ for the filtered field, in real space at $z=0$,
to the BBKS prediction.  All the results have been scaled by a constant
to agree at low-$k$, i.e.~any linear bias has been removed.
The thick solid line is the linear theory result, for comparison.
At low-$k$ linear bias is a good approximation to these results, with the
bias becoming more scale dependent at higher-$k$.  The second feature
remains in all of these schemes.}
\label{fig:bias}
\end{figure}

Two effects need to be considered further before a case can be made
for the detectability of baryonic features. The first is the question of
whether bias alters their visibility. We have assumed linear bias,
i.e.~$\delta N_g/ N_g = b\delta\rho_m/\rho_m$, where $N_g$ denotes the galaxy
counts, $\rho_m$ is the total matter density, and $b$ is the bias factor.
However, while semi-analytic modeling has made some progress in understanding
bias \cite{KauNusSte}, it is not a priori clear whether the
transformation between mass and light affects the relative visibility of the
baryonic features.
For the $\Lambda$CDM model we have carried out experiments with simple
nonlinear modifications of the final density field, such as
$\rho\rightarrow \rho^B$.
This confirms that, on sufficiently large scales, linear bias is indeed a
good approximation, even when the spectrum contains features
(see also Scherrer \& Weinberg~1998).
Some results for $\rho^B$ ``biasing'' are shown in Fig.~\ref{fig:bias}.
Note that on large scales the bias is linear, while it becomes more scale
dependent on small scales.
Even for large enough $B$ that the scale dependence would be strong,
the second peak in the power spectrum is still present.
These experiments remain preliminary. A definitive answer to this question
would require an understanding of the nature of galaxy bias.  As a first
step beyond our simple model one could compute the halo-halo correlation
function, or include hydrodynamics to identify sites of galaxy formation.
Neither of these approaches is feasible with our PM code.

Secondly, when comparing power spectra measured at different redshifts, care
must be taken to account for evolution.
Within the biased galaxy formation paradigm the evolution of
$\Delta^2_{\rm gal}$ with $z$ can be quite complicated and non-monotonic,
unlike the evolution of $\Delta^2_{\rm mass}$.
As has been pointed out recently by several authors
\cite{BraVil,OgaRouYam,Bag,PSFW} the contribution to the clustering strength
comes both from the underlying mass power spectrum (which grows with $z$)
and the bias of the object which is a function of its rarity at a given $z$
(and typically decreases with $z$).
Preliminary observational evidence for evolution has recently been
found \cite{LyBreak,ConSzaBru}.
Although eventually it may be possible to avoid the uncertainties introduced
by bias by measuring the dark matter fluctuations directly through
gravitational lensing \cite{Sel97,KaiLens}, for the upcoming large surveys it
will remain an issue.

Lastly, this analysis has focused on modes with $k\ga0.04\hompc$,
since on scales larger than this several systematic issues become important.
For example, a correction for extinction to 10\% of the true power across
the survey is required to control systematic errors on the power spectrum on
larger scales (for currently popular models), and this is quite difficult to
achieve.  Correlations in the photometric errors have similar effects.

\subsection{2D Surveys} \label{sec:2dsurvey}

\begin{figure}
\begin{center}
\leavevmode \epsfxsize=3.3in \epsfbox{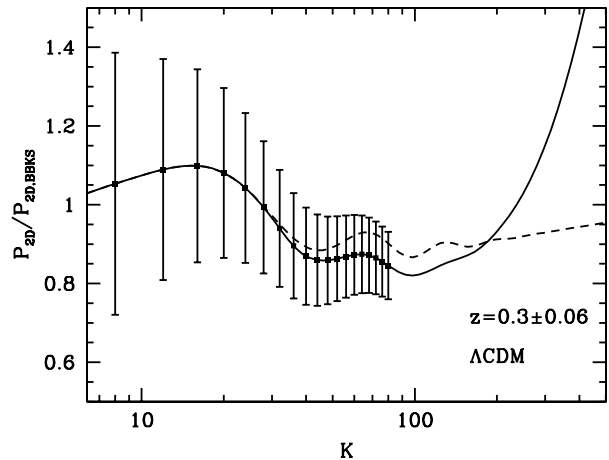}
\end{center}
\caption{The persistence of oscillations in the 2-dimensional power spectrum
$P_{2D}(K)$.  Solid lines show the full non-linear calculation, dashed lines
linear theory.  The galaxy selection functions is a Gaussian centered at
$z=0.3$ with width 20\%, which should be easily selectable using photometric
redshifts.  Note that the features, which become immeasurably small with an
APM-like selection function, survive convolution with a narrow redshift window.
The error bars approximate those expected from the SDSS survey. The model
is the same as in Figure \ref{fig:fitbar}.}
\label{fig:p2d}
\end{figure}

The discussion so far has focused exclusively on redshift surveys. However for
the SDSS there will be a much larger number of objects for which the photometry
and positions will be known, but not the redshifts.  As
mentioned previously, in a fully 2D survey the oscillations in the power
spectrum will become immeasurably small because of projection effects.  But by
making use of photometric redshifts, the photometric catalogue can be
used to select redshift shells, which may easily be $<20\%$ in width
\cite{BCSB}.  With such shells, the features in large volume 3D
surveys are {\it not\/} significantly washed out by projection.  We
shall work with shells of 20\% width.  Such relatively broad shells
have the advantage of containing a large number of galaxies, thus
reducing the shot noise. As it turns out, we will show that it is not
the shot noise that is the limiting factor, but rather the number of
uncorrelated measurements of the power spectrum that may be made.  For
this reason 3D surveys hold more promise than quasi-2D surveys for finding
the baryonic features.

A second reason to consider relatively broad shells is to reduce the effect of
redshift space distortions. Photometric redshifts are a combination of true
redshifts (since they are sensitive to the shifting of spectral breaks through
photometric filters), and distance (since fluxes drop off as $r^{-2}$).
If only colors are used to compute the redshift then the errors on $z$ are
larger, but the selection is more purely a redshift measurement. Selecting
galaxies on the basis of redshift enhances the resulting power spectrum over
the real space power spectrum \cite{Kaiser}. It is straightforward to show,
however, that if the selection function $dN_{\rm gal}/dz$ is slowly varying on
the scales $k^{-1}$ under consideration, one accurately
measures the real space power spectrum. More exactly, for a Gaussian shell
of width $\sigma$, the increase in the real space power spectrum
(or angular correlation function) to leading order in $k\sigma\gg1$ is
${\cal O}[f/(k\sigma)^2]$ or smaller, where $f(\Omega_0)\simeq\Omega_0^{0.6}$. 

To see this, compare the $k-$component of a planewave redshift space
fluctuation $\hat\delta_s({\bf k})$ with the real space fluctuation
$\hat\delta_r({\bf k})$:
$\hat\delta_s({\bf k})=(1+f\mu^2)\hat\delta_r({\bf k})$,
where ${\bf r}\cdot{\bf k}=rk\mu$.
The projected counts along a given line of sight are then
$\delta N=\int\,ds\;\delta n_s(r)\approx\int\,ds\;\delta_s(r)\bar\rho_s\phi(s)$,
where $\phi(s)$ is the selection function in redshift space.
Using the planewave decomposition and retaining terms of only first order in
$\delta$ gives
$\delta N\approx\int\, dk\; k^2 \int\, d\mu\; \bar\rho_r\hat\delta_r(k)
(1+f\mu^2)\int\, ds\; \phi(s)\exp(ik\mu s)$.
If $\phi(s)$ is a Gaussian of width $\sigma$, then
$\int\, ds\; \phi(s)\exp(ik\mu s)=\exp[-\sigma^2(k\mu)^2/ 2]$.
The integration over $\mu$ then ensures $\delta N$ agrees with the real space
projected counts to order $f/(k\sigma)^2$. A similar analysis starting with
Limber's equation yields an identical result.

On scales $k\sim0.1\hompc$ this correction is negligible for the
20\% width we have chosen.
If we probed larger scales, $k\to0$, we would eventually recover the famous
Kaiser factor $(1+2f/3+f^2/5)$, which enhances large-scale power in redshift
space.  For 20\% shells this enhancement could affect the first feature in
$P(k)$ near $k\sim0.02\hompc$,
and a more sophisticated analysis would be needed.

We show a 2D projection power spectrum in Fig.~\ref{fig:p2d}.
The circular Sloan area has an effective radius of $\Theta=1$~radian
(for a flat sky).  Its transform is $2J_1(K\Theta)/(K\Theta)$, which has
its first zero at $K=3.83/\Theta$.
We should therefore be able to deduce uncorrelated estimates of angular
power spectra, provided the bins in angular wavenumber are wider than
$\Delta K=4$.
The expected fractional power errors in this case are
\begin{equation}
  {\sigma_P\over P}=\sqrt{4\pi\over A K\Delta K} = (K/1.0)^{-1/2},
\end{equation}
assuming the minimal uncorrelated sampling of $\Delta K=4$.
Note that, in 2D, the error bars decline with wavenumber much more slowly
than in 3D, reflecting the larger number of modes in the 3D case.
This difference means that it is much harder to detect the baryon oscillations
in the 2D case; for the models considered here, the characteristic second
harmonic would never be detectable.
Since shot noise is not a major contributor to the error bars in either 2D
or 3D, the advantage of a larger number of objects in the sample is vastly
outweighed by the reduced number of modes.
This consideration obviously applies to other methods of measuring the
power spectrum (e.g.~lensing) where only 2D information is available.

\section{Summary} \label{sec:summary}

If the baryonic density is an appreciable fraction of the matter density
then the large-scale power spectrum is predicted to have features in addition
to the break at the scale of matter-radiation equality.
We have examined the observability of such features, specifically the
oscillations on scales of 10--$100\mpcoh$ which are the remnants
of acoustic oscillations in the photon-baryon fluid before last scattering.
Our principal conclusions may be summarized as follows:

(1) The linear theory of these processes is well understood, and may
be described as resulting in a series of `bumps' in power relative
to a reference zero-baryon model.

(2) The effect of largest amplitude (a factor perhaps 1.5 in power)
is a broad hump at low $k$. A clearer observational feature is
the second harmonic, which is expected to lie near $k=0.055\,\rm Mpc^{-1}$
for most models, and which is relatively narrow in $k$ space.

(3) Non-linear effects act to wash out the oscillations at higher $k$ due to
mode coupling.  Second order perturbation theory provides a qualitative
description of the effect, but is not quantitatively accurate.

(4) For acceptable values of the cosmological and baryon densities,
the oscillations may be measurable with forthcoming surveys. This
will best be done with three-dimensional surveys, rather than
quasi-2D surveys based on photometric redshifts or lensing.
The measurement of a baryon signature, however, will be extremely demanding.
A detection of the broad low-$k$ hump, the clearest signature of the
baryons, will require accurate measurements of the power spectrum
for $k<0.05\, h\, {\rm Mpc}^{-1}$. Because of degeneracy between models,
an unambiguous detection of the second harmonic will require a precise
determination of the amplitude of the fluctuations, as could be provided
by measurements of the CMB anisotropy for example.

\bigskip
M.W. would like to thank Matt Craig, Marc Davis and George Efstathiou for
useful conversations on N-body codes and Wayne Hu for conversations on
baryonic features.  A.M. and M.W. thank Alex Szalay for many useful
conversations on a variety of topics. The authors thank the anonymous
referee for a very careful reading of the manuscript and helpful comments.
The results in this paper made use of the Hydra $N$-body code \cite{Hydra}.
A.M. and M.W. thank Hugh Couchman for enlightenment on the operation of Hydra.
M.W. was supported by the NSF.

\appendix

\section{First-order perturbations} \label{sec:theory}

The matter and radiation power spectra used in this paper were calculated
numerically by evolution of the coupled Einstein, fluid and Boltzmann equations
in synchronous gauge (for details of the code see
White \& Scott 1996; Hu et al. 1996).
However to understand the physics behind the oscillations it is simpler to
use the Newtonian gauge.  The relation between these two gauges is discussed
in e.g.~Hu, Spergel \& White~\shortcite{HSW}.

We start with the fundamental equations describing the dynamics of a
relativistic fluid.  A physical interpretation and description of these
equations can be found in Hu \& White~(1996), to which the reader
is referred for more details.
The evolution of the photons and baryons in a metric perturbed
by density fluctuations in the $k$th normal mode is given in the Newtonian
gauge as
\begin{equation}
\begin{array}{rcl}
\dot{\Theta}_0 &=& -{\displaystyle{k \over 3}} \Theta_1 - \dot\Phi, \\
\dot{\delta}_b &=& -kV_b - 3\dot\Phi,
\end{array}
\end{equation}
for the continuity equations and
\begin{equation}
\begin{array}{rcl}
\dot{\Theta}_1 &=& k[\Theta_0 + \Psi - {1 \over 6}\Pi_\gamma] -
        \dot \tau(\Theta_1 - V_b), \\
\dot{V}_b &=& - {\displaystyle{\dot a \over a}}V_b + k\Psi +
        \dot \tau(\Theta_1 - V_b)/R,
\end{array}
\label{eqn:Euler}
\end{equation}
for the momentum conservation or Euler equations of the photons and baryons
respectively.  The evolution of the cold dark matter is the same as the
baryons except for the $\dot\tau$ terms which describe the coupling to the
photons.
Overdots are derivatives with respect to conformal time $\eta=\int dt/a$,
$R\equiv 3\rho_b/4\rho_\gamma$ is the baryon-photon momentum density ratio,
and $\dot{\tau}$ is the differential Compton optical depth.
The fluctuations are defined as $\Theta_0 =\Delta T/T=\delta_\gamma/4$
the isotropic temperature perturbation, $\Theta_1$ the dipole moment
or photon bulk velocity, $\Pi_\gamma$ the photon anisotropic stress
perturbation, $\delta_b$ the baryon energy density perturbation and
$V_b$ the baryon velocity.  The gravitational sources are $\Phi$,
the perturbation to the spatial curvature, and $\Psi$, the Newtonian
potential.  At late times these are dominated by the CDM and baryons (if
$\Ob/\Omega_0$ is large) while at early times they are dominated by the
relativistic species.
The Einstein equations are
\begin{equation}
\begin{array}{rcl}
(k^2 - 3K) \Phi &=& 4\pi G a^2 \sum \left[ \rho_i \delta_i + 3
\displaystyle{\dot a \over a}
(\rho_i + p_i)V_i/k \right], \\
k^2 (\Psi + \Phi) &=& -8\pi G a^2 \sum p_i \Pi_i,
\end{array}
\label{eqn:Poisson}
\end{equation}
where the sum is over particle species and the curvature
$K=-H_0^2(1-\Omega_0-\Omega_\Lambda)$.

At early times the density is high and the scattering is rapid compared with
the travel time across a wavelength.  Thus we may expand the momentum
conservation equation in powers of the Compton mean free path over the
wavelength $k/\dot{\tau}$.  To lowest order we obtain the
{\it tight coupling\/} approximation for the evolution
\begin{equation}
{d \over d\eta}(1+R)\dot\delta_b + {k^2 \over 3}\delta_b
  = -k^2(1+R)\Psi - {d \over d\eta}3(1+R)\dot\Phi.
\end{equation}
which is a driven harmonic oscillator with natural frequency
$c_s^{-2}=3(1+R)$.  During the tight coupling phase the amplitude of the
baryon perturbations cannot grow, it undergoes harmonic motion with an
amplitude which decays as $(1+R)^{-1/4}$ and a velocity which decays as
$(1+R)^{-3/4}$.  For values of $\Ob$ consistent with standard BBN, $R<0.3$
so this effect is not dominant.
If we define the optical depth
$\tau_b(\eta)\equiv\int_\eta^{\eta_0}\ \dot\tau d\eta'/(1+R)$, we find that
the baryons decouple from the photons when $\tau_b\sim1$.
The oscillations in the baryons are frozen in at this epoch.
Expanding to higher order in $k/\dot{\tau}$ one finds the oscillations are
exponentially damped with characterstic scale (for more details and a
discussion of the physics of the damping see e.g. Hu \& White~1997)
\begin{equation}
k_D^{-2}(\eta) = {1 \over 6} \int d\eta\ {1\over\dot{\tau}}\,
{R^2 + 16(1+R)/15 \over (1+R)^2}
\end{equation}
where $k_D(\eta)$ is evaluated at the peak of the visibility function
$\dot\tau_b\exp(-\tau_b)$.
For $\Ob$ consistent with standard BBN this decoupling is after
recombination.  In any case, this damping turns out not to be
phenomenologically interesting, since non-linear effects wash out the
higher peaks anyway.

Once the photons release their hold on the baryons we can neglect the
$\dot{\tau}$ terms in Eq.~(\ref{eqn:Euler}).  The solutions are then the
well known growing and decaying modes of pressureless linear perturbation
theory (e.g.~in a critical density, matter dominated universe the growing
mode $\propto a$).  The density and velocity perturbations from the
tight-coupling era must be matched onto the growing and decaying modes 
for the pressureless components, including the CDM potentials.  The final
spectrum is the component which projects onto the growing mode.
As is well known (e.g., Padmanabhan 1993, \S8.2) at high $k$ the growing
mode is sourced primarily by the velocities, while at low-$k$ it is a
mixture of density and velocity terms.  For this reason at high $k$ the
oscillations in $\Delta_{\rm L}^2(k)$ are out of phase with the peaks in the
CMB anisotropy spectrum which arise predominantly from photon densities.
In this limit peaks occur at $kr_*=(2j+1)\pi/2$ where $j=0,1,2,\cdots$ and
$r_*$ is the sound horizon at decoupling: $r_*=\int c_s\, d\eta$.  In detail
this differs from the sound horizon at recombination (which controls the
position of the CMB peaks) but as can be seen in Fig.~\ref{fig:cmblss} the
two horizons are comparable.

In Fig.~\ref{fig:cmblss} we see that the baryons induce both oscillations and
amplitude suppression in $T(k)$.  Let us consider the latter first.
Modes which enter the horizon before matter-radiation equality can grow only
logarithmically at best.  For the baryons, modes which enter the horizon
{\it after\/} equality but before decoupling oscillate with decaying amplitude.
They therefore do not contribute fully to the gravitational potentials
in Eq.~(\ref{eqn:Poisson}).
For the CDM the modes which enter the horizon after equality but before
decoupling grow as if in a universe with density $\approx\Omega_0-\Ob$ rather
than $\Omega_0$.  Hence for the period between equality and decoupling all
modes which are inside the horizon have their growth suppressed relative to
the $\Ob\ll\Omega_0$ case.  This damping of power changes the shape of
$P(k)$ near its peak, with the largest effect occuring in models where
equality and decoupling are most separated
(high $\Ob h^2$ and high $\Omega_0 h^2$).

The acoustic oscillations in the baryons at decoupling \cite{HuSug}
are then superposed upon the smooth power spectrum. The amplitude of the
oscillations depends on both the driving force ($\Phi$ and $\Psi$) and on $R$.
Since the potentials decay in the radiation dominated epoch, larger
oscillations come from higher $\Ob h^2$ and lower $\Omega_0 h^2$
(see e.g.~discussion of the potential envelope in
Hu \& White 1997).
Specific examples are shown in Fig.~\ref{fig:del2}.

\section{PM Code} \label{sec:pmcode}

To investigate the effects of non-linearity on the persistence of the
features in the power spectrum we used a PM code to calculate the non-linear
power spectrum.  Since it has not been discussed before, we give some details
of our implimentation of the code here.  All of our results were also checked
by running P$^3$M simulations as discussed in the text.  Above the grid scale
the agreement was excellent.

We used $128^3$ particles with the gravitational forces computed on a $128^3$
grid.  All of the runs were started at $1+z=20$, and evolved using equations
of motion in which the `time' coordinate is $\ln a$, where $a$ is the scale
factor.
The initial particle positions were displaced from a random position within a
cell (with cells uniformly filling the box) using the Zel'dovich approximation.
Each realization of the power spectrum was chosen to have random phases and
amplitude drawn from an exponential distribution.

To calculate the forces on the particles we assigned them to a grid using the
cloud-in-cell algorithm.  On large scales the power spectrum recovered is
independent of the charge assignment scheme, which we explicitly checked.
The forces were computed using FFT techniques with a force kernel of
$\vec{k}/k^2$ to compute $\vec{\nabla}\Phi$.  While it is computationally
faster to compute $\Phi$ directly using a $1/k^2$ kernel and calculate
$\vec{\nabla}\Phi$ by differencing, we found that this gave worse performance
on small scales than calculating $\vec{\nabla}\Phi$ for each direction using
$\vec{k}/k^2$.  The better accuracy at small scales in this method more than
makes up for the extra computing time (compared to running higher resolution).
The time step was dynamically chosen as a small fraction of the inverse square
root of the maximum acceleration, with an upper limit of $\Delta a/a=4\%$ per
step, where $a$ is the cosmological expansion factor. This resulted in a final
particle position error of less than 0.1\% of the box size.

At selected time steps the power spectrum was calculated.
The density field $\delta_k$ was computed by assigning all particles to the
nearest node in a $128^3$ grid and performing an FFT.
The resulting power, $|\delta_k|^2$, was corrected for shot-noise and binning
onto the grid to obtain an estimate for $P(k)$.
The redshift space power spectrum was computed by adding one component of the
velocity (in units of the Hubble constant times the length of the box) to the
particle positions before assigning them to the grid.

\section{Scale--Dependent Bias} \label{sec:sdbias}

Let us imagine that we have two classes of objects, whose probability of being
included in the sample at a position ${\bf R}$, is $\phi_a({\bf R})$ with
$a=1,2$.
(Luminosity is only one possible criterion for inclusion in a catalog; surface
brightness could be another.) A straightforward calculation gives
\begin{equation}
\xi_{ij} = 
  { \phi_{1i}\phi_{1j}\xi_{11} +
    (\phi_{1i}\phi_{2j}+\phi_{2i}\phi_{ij})\xi_{12} +
    \phi_{2i}\phi_{2j}\xi_{22} \over
    (\phi_{1i}+\phi_{2i}) (\phi_{1j}+\phi_{2j}) }
\end{equation}
where $i,j$ label positions in the survey and we have used the shorthand
$\phi_{1j}=\phi_1({\bf R}_j)$.
If we further suppose that $\xi_{11}(r)=b_1^2\xi(r)$,
$\xi_{22}(r)=b_2^2\xi(r)$ and $\xi_{12}(r)=b_1 b_2\xi(r)$, where $\xi(r)$
represents the correlation function for points at a separation $r$ drawn from
the underlying field, of which both populations are biased tracers.
We find
\begin{eqnarray}
b^2(R, r) &\equiv& \left\langle{\xi^{\rm est}({\bf R}, r)\over \xi(r)}\right
\rangle \\
&=& N_{\rm pr}^{-1} \sum_{ij}
  { (b_1\phi_1+b_2\phi_2)_i (b_1\phi_1+b_2\phi_2)_j \over
    (   \phi_1+   \phi_2)_i (   \phi_1+   \phi_2)_j }
\end{eqnarray}
where $R$ is the depth of the survey, the averaging is over the survey area,
the sum is over all positions $i,j$ in the survey which are separated by a
distance $r$ and $N_{\rm pr}$ is the number of such pairs of positions. The
resulting power spectrum will be
\begin{equation}
\Delta^2(R,k)={1\over{(2\pi)^3}}\int\,d^3q\,\hat{b^2}
(R,|{\bf k}-{\bf q}|)\Delta_0^2(R,q)
\end{equation}
where $\hat{b^2}(R,k)$ is the Fourier transform of $b^2(R,r)$ and
$\Delta_0^2(k)$ is the power spectrum assuming no bias.

If the selection functions vary smoothly throughout the survey volume, the
inferred bias parameter will depend only on the survey depth $R$ as the
ratio $\phi_1(R)/ \phi_2(R)$ varies with $R$. This results in an overall shift
with depth of the amplitude of the measured
power spectrum, which would need to be distinguished from
evolution in the clustering. It is also possible, however, to introduce a
feature into the power spectrum at a given depth if the selection function of
one of the populations changes rapidly in the survey volume. This could occur
in a redshift slice if the selection function of one population either was
rapidly declining in the slice or occupied a very narrow range in redshift
space. For galaxies at a fixed mean redshift in a very narrow redshift shell, 
the population composition that dominates the contribution to the power
spectrum at a given scale will differ for different scales. Depending on the
shapes and peaks of the selection functions, and the position of the shell,
this can result in either enhancements or reductions in the power spectrum
with increasing scale.

\begin{figure}
\begin{center}
\leavevmode \epsfxsize=3.3in \epsfbox{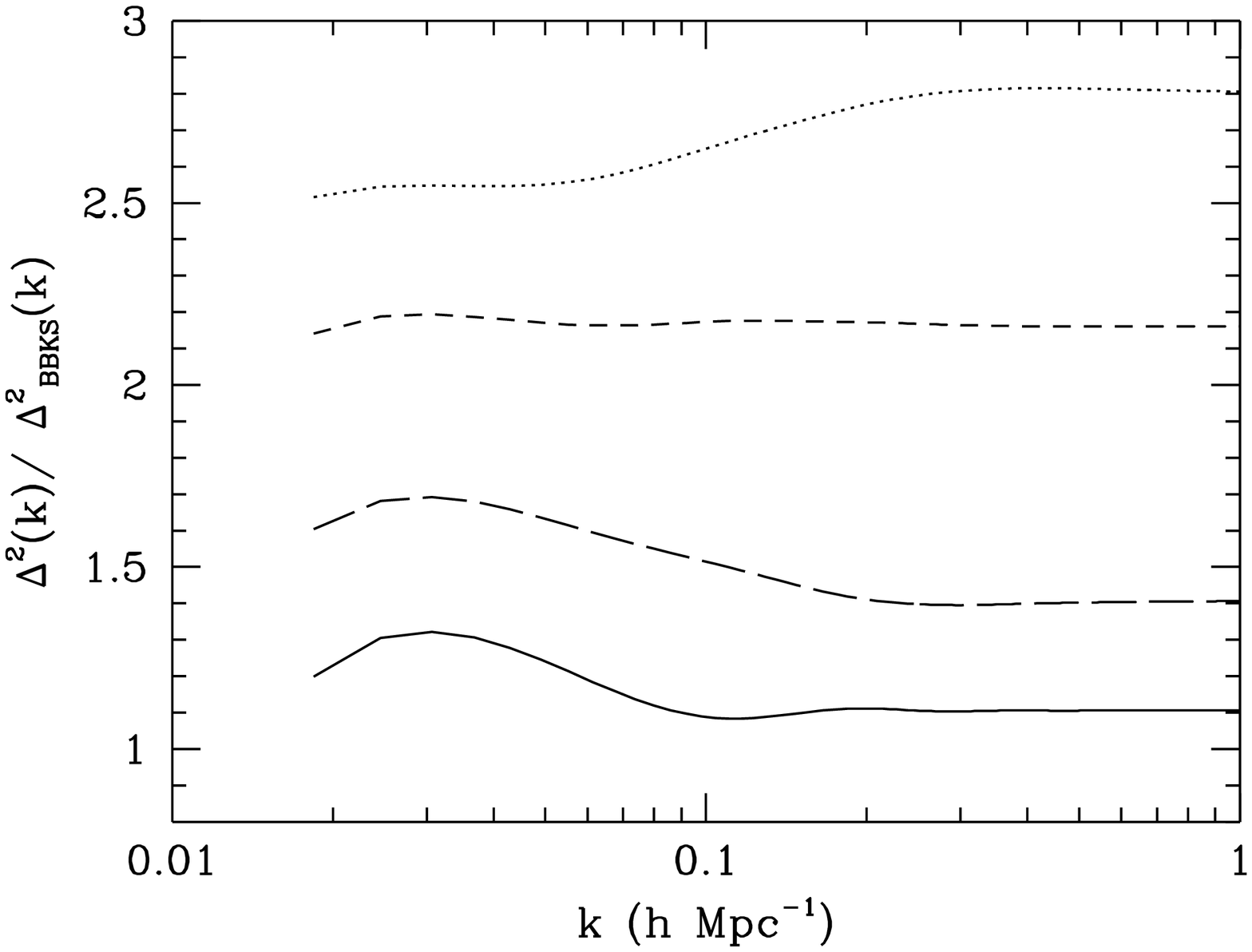}
\end{center}
\caption{
Scale--dependent bias is introduced by mixing a dominant population having a
broad selection function peaking at $z_1=0.2$ and bias $b_1=2$, with
a minority unbiased ($b_2=1$) population having a selection function that is
sharply peaked at $z_2=0.3$. The resulting distortion
of the assumed underlying BBKS power spectrum is shown for redshift shells
positioned at $z=0.3$ ({\it solid}), $z=0.295$ ({\it long--dashed}), $z=0.293$
({\it short--dashed}), and $z=0.292$ ({\it dotted}). The features could easily
be mistaken for acoustic oscillations at $z=0.3$, and would suppress the
oscillations at $z=0.292$. They would, however, be averaged away in a shell
of full width $\Delta z=0.1$.}
\label{fig:sdbias}
\end{figure}

We illustrate the effect in Figure \ref{fig:sdbias}. The survey volume is taken
to be cone with solid angle $\pi$ sr, as for the SDSS. The selection functions
for both populations are assumed to be Gaussians with mean redshift $\bar z_i$
and width $\sigma_i$. One population is taken to comprise 75\% of the galaxies
(integrated over all redshifts), with $\bar z_1=0.2$, $\sigma_1=0.1$, and
$b_1=2$. The selection function for the remaining population is very narrowly
peaked at $\bar z_2=0.3$ with $\sigma_2=0.003$. It is taken to be unbiased
($b_2=1$). The underlying matter power spectrum is assumed to be BBKS CDM.
For a shell centered at $z=0.3$, an oscillating scale--dependent
bias is produced, which could easily be mistaken for an acoustic oscillation.
The effect, however, diminishes sharply as the shell is moved away from
$z=\bar z_2$, until by $z=0.25$ the distinctions are reduced to less that 1\%.
This means that for a shell of full width $\Delta z=0.1$, the oscillations
would be completely masked.

The strength of the effect is entirely a consequence of the narrowness of
the selection function for the second population, which we consider to be
unphysical. Moreover, even if the effect were detected, its extreme sensitivity
to redshift would distinguish it from the acoustic oscillations, which are not
expected to change with redshift except by the much more gradual growth due to
gravitational instability. Typically, the effect on the power spectrum is much
smaller than that shown here, even in the presence of a substantial
scale--dependent bias. For instance, if the width of the second population
above were increased to $\sigma_2=0.05$, with everything else held unchanged,
at $z=0.3$ the distortions of the power spectrum would be only 0.1\%, even
though the mean bias factor changes by 3\% between 20 and $200\mpcoh$.

\section{Galaxy Number Counts} \label{sec:counts}

The number of galaxies per unit redshift $dN/dz$ to a limiting magnitude
$m_{\rm lim}$ in a survey of area $\Omega_s$ sr is related to the proper
luminosity function $\phi(M,z)$ of the galaxies by

\begin{equation}
{dN\over dz}(<m_{\rm lim})=\Omega_s d_{\rm A}^2(z)\ {d\ell_p\over dz}
\int_{-\infty}^{M_{\rm max}}\; dM\; \phi(M,z),
\end{equation}
where
\begin{equation}
  M_{\rm max}=m_{\rm lim}-5\log_{10}d_{\rm L}(z)-25-K(z)-A,
\end{equation}
$d_{\rm L}(z)$ is the luminosity distance (in Mpc) to a source at redshift
$z$, $d_{\rm A}(z)=d_{\rm L}(z)/ (1+z)^2$ is the angular distance to the
source, and $d\ell_p$ is the proper differential line element.
The terms $K(z)$ and $A$ account for the K--correction and extinction,
respectively.
For $\Lambda=0$,
\begin{equation}
  d_{\rm L}(z)={2c\over H_0\Omega_0^2}\left\{z\Omega_0-(2-\Omega_0)
  \left[ (1+\Omega_0 z)^{1/2}-1\right]\right\},
\end{equation}
and
\begin{equation}
{d\ell_p\over dz}={c\over H_0}(1+z)^{-2}(1+\Omega_0z)^{-1/2}.
\end{equation}
When $\Lambda\ne0$ closed form expressions do not exist and the defining
integrals need to be evaluated numerically.

For a Schechter luminosity function, $\log_{10}(L/L_{*})=0.4(M^*-M)$ and
\begin{equation}
\begin{array}{lcl}
  \phi(M,z) &=& (0.4\log10)\ (1+z)^3\ \phi^*(z) \\
  && \times (L/L_*)^{1+\alpha} \exp[-L/L_*],
\end{array}
\end{equation}
where the density expansion rate $(1+z)^3$ has been factored out so that
any residual redshift dependence in $\phi^*(z)$ is due to galactic evolution.
The integral over $M$ is now analytic.
Defining $x_{\rm min}$ by $2.5\log_{10}x_{\rm min}=M^*-M_{\rm max}$,
\begin{eqnarray}
\int_{-\infty}^{M_{\rm max}}\; dM\; \phi(M,z)=(1+z)^3\phi^*(z)
\Gamma(1+\alpha,x_{\rm min}),
\end{eqnarray}
where $\Gamma(\alpha,x)$ is an incomplete gamma function
\cite{GraRyz}. For $-2<\alpha<-1$, it may be evaluated using
the recursion relation,
\begin{equation}
\Gamma(1+\alpha,x)=\frac{1}{1+\alpha}\left[\Gamma(2+\alpha,x)
-x^{1+\alpha}e^{-x}\right].
\end{equation}

We have used these estimates to obtain the number of galaxies in redshift
shells in \S\ref{sec:2dsurvey} and hence the shot-noise contribution to the
error bar in Fig.~\ref{fig:p2d}.

\label{lastpage}


\begin{thebibliography}{99}
\bibitem[\protect\citename{Amendola }1994]{Amen94}
Amendola L., 1994, ApJ, 430, L9
\bibitem[\protect\citename{Bagla }1998]{Bag}
Bagla J.S., 1998, MNRAS, 299, 417
\bibitem[\protect\citename{Bardeen, Bond \& Efstathiou }1987]{BBE}
Bardeen J.M., Bond J.R., Efstathiou G., 1987, ApJ, 321, 28
\bibitem[\protect\citename{Bardeen et al. }1986]{BBKS}
Bardeen J.M., Bond J.R., Kaiser N., Szalay A.S., 1986, ApJ, 304, 15
\bibitem[\protect\citename{Blumenthal et al. }1988]{BDP}
Blumenthal G.R., Dekel A., Primack J.R., 1988, ApJ, 326, 539
\bibitem[\protect\citename{Brainerd \& Villumsen }1994]{BraVil}
Brainerd T., Villumsen J.V., 1994, ApJ, 431, 477
\bibitem[\protect\citename{Broadhurst et al. }1990]{BEKS}
Broadhurst T.J., Ellis R.S., Koo D.C., Szalay A.S., 1990, Nature, 343, 726
\bibitem[\protect\citename{Bromley \& Press }1998]{BroPre}
Bromley B., Press W.H., in preparation.
\bibitem[\protect\citename{Bromley et al. }1998]{BPLK}
Bromley B., Press W.H., Lin H., Kirshner R.P., preprint, astro-ph/9805197
\bibitem[\protect\citename{Brunner et al. }1998]{BCSB}
Brunner R.J., Connolly A.J., Szalay A.S. \& Bershady M.A., 1998,
  ApJ, 482, L21
\bibitem[\protect\citename{Bunn \& White }1997]{BunWhi}
Bunn E.F. \& White M., 1997, ApJ, 480, 6
\bibitem[\protect\citename{Catelan et al. }1995]{Cat95}
Catelan P., Lucchin F., Sabino S., Moscardini L., 1995, MNRAS, 276, 39
\bibitem[\protect\citename{Cole, Fisher \& Weinberg }1995]{ColFisWei}
Cole S., Fisher K.B., Weinberg D.H., 1995, MNRAS, 275, 515
\bibitem[\protect\citename{Connolly et al. }1995]{Con95}
Connolly A.J. et al., 1995, AJ, 110, 2655
\bibitem[\protect\citename{Connolly, Szalay \& Brunner }1998]{ConSzaBru}
Connolly A.J., Szalay A., Brunner R.J., 1998, ApJ, 499, L125
\bibitem[\protect\citename{Couchman }1991]{Cou}
Couchman H.M.P., 1991, ApJ, 368, L23
\bibitem[\protect\citename{Couchman, Thomas \& Pearce }1995]{Hydra}
Couchman H.M.P., Thomas P.A., Pearce F.R., 1995, ApJ, 452 797
\bibitem[\protect\citename{Davis et al. }1988]{dms}
Davis M., Meiksin A., Strauss M.A., da Costa L.N., Yahil A.,
1988, ApJ, 333, L1
\bibitem[\protect\citename{Dekel }1984]{Dek}
Dekel A., 1984, ApJ, 284, 445
\bibitem[\protect\citename{Einasto et al. }1997a]{EinA}
Einasto J., et al., 1997a, Nature, 385, 139
\bibitem[\protect\citename{Einasto et al. }1997b]{EinB}
Einasto J., et al., 1997b, MNRAS, 289, 801
\bibitem[\protect\citename{Eisenstein \& Hu }1998]{EisHu}
Eisenstein D.J., Hu W., 1998, ApJ, 496, 605
\bibitem[\protect\citename{Eisenstein et al. }1998]{EHSS}
Eisenstein D.J., Hu W., Silk J., Szalay A.S., 1998,
ApJ, 494, L1
\bibitem[\protect\citename{Eisenstein, Hu, \& Tegmark}1998]{EHT}
Eisenstein D.J., Hu W., Tegmark M. 1998, preprint, astro-ph/9807130
\bibitem[\protect\citename{Eke, Cole \& Frenk }1996]{Eke96}
Eke V.R., Cole S., Frenk C.S., 1996, MNRAS, 282, 263
\bibitem[\protect\citename{Eke, Cole, Frenk \& Henry }1998]{Eke98}
Eke V.R., Cole S., Frenk C.S., Henry J.P., 1998, MNRAS, 298, 1145
\bibitem[\protect\citename{Elbaz, Arnaud \& B{\"o}hringer }1995]{ElbArnBoh}
Elbaz D., Arnaud M., B{\"o}hringer H., 1995, A\&A, 293, 337
\bibitem[\protect\citename{Evrard, Metzler \& Navarro }1996]{EvrMetNav}
Evrard A.E., Metzler C., Navarro J.F., 1996, ApJ, 469, 494
\bibitem[\protect\citename{Fan \& Bardeen }1995]{FB95}
Fan Z., Bardeen J.M., 1995, PRD, 51, 6714
\bibitem[\protect\citename{Feldman, Kaiser \& Peacock }FKP]{FKP}
Feldman H.A., Kaiser N. \& Peacock J.A., 1994, ApJ, 426, 23
\bibitem[\protect\citename{Giavalisco et al. }1998]{LyBreak}
Giavalisco M., et al., 1998, ApJ, 503, 543
\bibitem[\protect\citename{Goldberg \& Strauss }1998]{GolStr}
Goldberg  D.M., Strauss M., 1998, ApJ, 495, 29
\bibitem[\protect\citename{Gradshteyn \& Ryzhik }1980]{GraRyz}
Gradshteyn I.S., Ryzhik I.M., 1980, Table of Integrals, Series and
Products. Academic Press, New York
\bibitem[\protect\citename{Gunn \& Weinberg }1995]{GW}
Gunn J., Weinberg D., 1995, in Maddox S.~J., Arag\'on-Salamanca A., eds,
Wide-field spectroscopy and the Distant Universe. World Scientific, p.3
\bibitem[\protect\citename{Gyuk \& Turner }1994]{GuyTur}
Gyuk G., Turner M.S., 1994, Phys. Rev. D, 50, 6130
\bibitem[\protect\citename{Hu et al. }1995]{HSSW}
Hu W., Scott D., Sugiyama N., White M., 1995, PRD, 52, 5498
\bibitem[\protect\citename{Hu, Spergel \& White }1997]{HSW}
Hu W., Spergel D., White M., 1997, PRD, 55, 3288
\bibitem[\protect\citename{Hu \& Sugiyama }1996]{HuSug}
Hu W., Sugiyama N., 1996, ApJ, 471, 542
\bibitem[\protect\citename{Hu \& White }1996]{HuWhi}
Hu W., White M., 1996, ApJ, 471, 30
\bibitem[\protect\citename{Hu \& White }1997]{Damping}
Hu W., White M., 1997, ApJ, 479, 568
\bibitem[\protect\citename{Jain \& Bertschinger }1994]{Jain94}
Jain B., Bertschinger E., 1994, ApJ, 431, 495
\bibitem[\protect\citename{Juszkiewicz }1981]{Jusz81}
Juszkiewicz R., 1981, MNRAS, 197, 931
\bibitem[\protect\citename{Kaiser }1987]{Kaiser}
Kaiser N., 1987, MNRAS, 227, 1
\bibitem[\protect\citename{Kaiser }1998]{KaiLens}
Kaiser N., 1998, ApJ, 498, 26
\bibitem[\protect\citename{Kaiser \& Peacock }1991]{KP91}
Kaiser N., Peacock J.A., 1991, ApJ, 379, 482
\bibitem[\protect\citename{Kauffmann, Nusser \& Steinmetz }1997]{KauNusSte}
Kauffmann G., Nusser A., Steinmetz M., 1997, MNRAS, 286, 795
\bibitem[\protect\citename{Kenney \& Keeping }1959]{KK59}
Kenney J.F., Keeping E.S., 1959,
  Mathematics of Statistics. D. van Nostrand Co., Princeton
\bibitem[\protect\citename{Landy et al. }1996]{Lanetal}
Landy S.D., Shectman S.A., Lin H., Kirshner R.P., Oemler A.A.,
  Tucker D., 1996, ApJ, 456, L1
\bibitem[\protect\citename{Lin et al. }1996]{LCRS}
Lin H., et al., 1996, ApJ, 471, 617
\bibitem[\protect\citename{Linsky et al. }1995]{Linetal}
Linsky J.L., Diplas A., Wood B.E., Brown A.,
  Ayres T.R., Savage B.D., 1995, ApJ, 451, 335
\bibitem[\protect\citename{Maddox, Efstathiou \& Sutherland }1996]{Mad96}
Maddox S.J., Efstathiou G., Sutherland W.J., 1996, MNRAS, 283, 1227
\bibitem[\protect\citename{Makino et al. }1992]{Mak92}
Makino N., Sasaki M., Suto Y., 1992, PRD, 46, 585
\bibitem[\protect\citename{Markevitch et al. }1996]{Maretal}
Markevitch R., Mushotzky R., Inoue H., Yamashita K., Furuzawa A., Tawara Y.,
  1996, ApJ, 456, 437
\bibitem[\protect\citename{Mathews, Kajino \& Orito }1996]{MKO}
Mathews G.J., Kajino T., Orito M., 1996, ApJ, 456, 98
\bibitem[\protect\citename{Meiksin \& White }1998]{MeiWhi}
Meiksin A., White M., 1998, preprint
\bibitem[\protect\citename{Ogawa, Roukema \& Yamashita }1997]{OgaRouYam}
Ogawa T., Roukema B.F., Yamashita K., 1997, ApJ, 484, 53
\bibitem[\protect\citename{Padmanabhan }1993]{Pad}
Padmanabhan T., 1993, Structure Formation in the
  Universe. Cambridge University Press, New York
\bibitem[\protect\citename{Peacock \& Dodds }1996]{PD96}
Peacock J.A., Dodds S.J., 1996, MNRAS, 280, L19
\bibitem[\protect\citename{Peacock }1997]{Pea}
Peacock J.A., 1997, MNRAS, 284, 885
\bibitem[\protect\citename{Peebles }1987a]{PeeA}
Peebles P.J.E., 1987a, ApJ, 315, L73
\bibitem[\protect\citename{Peebles }1987b]{PeeB}
Peebles P.J.E., 1987b, Nature, 327, 210
\bibitem[\protect\citename{Pen }1998]{Pen}
Pen U.-L., 1998, ApJ, 498, 60
\bibitem[\protect\citename{Press \& Vishniac }1980]{PreVis}
Press W.H. \& Vishniac E.T., 1980, ApJ, 236, 323
\bibitem[\protect\citename{Primack et al. }1998]{PSFW}
Primack J.R. et al., 1998, preprint, astro-ph/9806263
\bibitem[\protect\citename{Saunders, Rowan-Robinson \& Lawrence }1992]{Saun92}
Saunders W., Rowan-Robinson M., Lawrence A., 1992, MNRAS, 258, 134
\bibitem[\protect\citename{Scherrer \& Weinberg }1998]{SchWei}
Scherrer R.J., Weinberg D., 1998, ApJ, 504, 607
\bibitem[\protect\citename{Seljak }1997]{Sel97}
Seljak U., 1997, preprint, astro-ph/9711124
\bibitem[\protect\citename{Smith, Kawano \& Malaney }1993]{Smietal}
Smith M.S., Kawano L.H., Malaney R.A., 1993, ApJS, 85, 219
\bibitem[\protect\citename{Steigman \& Felten }1995]{SteFel}
Steigman G., Felten J.E., 1995, Spa. Sci. Rev., 74, 245
\bibitem[\protect\citename{Sugiyama }1995]{Sug}
Sugiyama N. 1995, ApJS, 100, 281
\bibitem[\protect\citename{Sunyaev \& Zel'dovich} 1970]{SunZel}
Sunyaev R.A., Zel'dovich Ya.B., 1970, Ap\&SS, 7, 3
\bibitem[\protect\citename{Tegmark }1997]{Teg97}
Tegmark M., 1997, Phys. Rev. Lett., 79, 3806
\bibitem[\protect\citename{Tegmark et al. }1998]{THSVS}
Tegmark M. et al., 1998, ApJ, 499, 555
\bibitem[\protect\citename{Tytler, Fan \& Burles }1996]{TytFanBur}
Tytler D., Fan X.-M., Burles S., 1996, Nature, 381, 207
\bibitem[\protect\citename{Viana \& Liddle }1996]{VL96}
Viana P.T.P., Liddle A.R., 1996, MNRAS, 281, 323
\bibitem[\protect\citename{Viana \& Liddle }1998]{VL98}
Viana P.T.P., Liddle A.R., 1998, preprint, astro-ph/9803244
\bibitem[\protect\citename{Vishniac }1983]{Vish83}
Vishniac E.T., 1983, MNRAS, 203, 345
\bibitem[\protect\citename{Walker et al. }1991]{Waletal}
Walker P.N., Steigman G., Schramm D.N., Olive K.A., Kang H.-S.,
  1991, ApJ, 376, 51
\bibitem[\protect\citename{Warren et al. }1994]{WHO94}
Warren S.J., Hewett P.C., Osmer P.S., 1994, ApJ, 421, 412
\bibitem[\protect\citename{White \& Fabian }1995]{WhiFab}
White D.A., Fabian A.C., 1995, MNRAS, 273, 72
\bibitem[\protect\citename{White, Efstathiou \& Frenk }1993a]{WEF}
White S.D.M., Efstathiou G.P., Frenk C.S., 1993a, MNRAS, 262, 1023
\bibitem[\protect\citename{White et al. }1993b]{Whietal}
White S.D.M., Navarro J.F., Evrard A.E., Frenk C.S., 1993b, Nature, 366, 429
\bibitem[\protect\citename{White \& Scott }1996]{WhiSco}
White M., Scott D., 1996, ApJ, 459, 415
\bibitem[\protect\citename{White \& Silk }1996]{WhiSil}
White M., Silk J., 1996, Phys. Rev. Lett., 77, 4704; erratum, 78, 3799
\bibitem[\protect\citename{White et al. }1996]{WVLS}
White M., Viana P.T.P., Liddle A.R., Scott D., 1996, MNRAS, 283, 107
\bibitem[\protect\citename{Willmer et al. }1998]{WDP}
Willmer C.N.A., daCosta L.N., Pellegrini P.S., 1998, AJ, 115, 869
\end{thebibliography}
\end{document}